# Analysis of Two-State Folding Using Parabolic Approximation I: Hypothesis

**AUTHOR NAME:** Robert S. Sade

**AUTHOR ADDRESS:** Vinkensteynstraat 128, 2562 TV, Den Haag, Netherlands

**AUTHOR EMAIL ADDRESS:** robert.sade@gmail.com

**AUTHOR AFFILIATION:** Independent Researcher

**KEY WORDS:** Arrhenius rate law; Marcus theory; parabolic approximation; transition state theory; two-state folding.

# ABSTRACT

A model which treats the denatured and native conformers of spontaneously-folding fixed two-state systems as being confined to harmonic Gibbs energy-wells has been developed. Within the assumptions of this model the Gibbs energy functions of the denatured (DSE) and the native state (NSE) ensembles are described by parabolas, with the mean length of the reaction coordinate (RC) being given by the temperature-invariant denaturant $m$ value. Consequently, the ensemble-averaged position of the transition state ensemble (TSE) along the RC, and the ensemble-averaged Gibbs energy of the TSE are determined by the intersection of the DSE and the NSE-parabolas. The equations derived enable equilibrium stability and the rate constants to be rationalized in terms of the mean and the variance of the Gaussian distribution of the solvent accessible surface area of the conformers in the DSE and the NSE. The implications of this model for protein folding are discussed.

# INTRODUCTION

Understanding the mechanism(s) by which denatured or nascent polypeptides under folding conditions spontaneously fold to their unique three-dimensional structures is one of the fundamental problems in biology. Although there has been tremendous progress since the ground-breaking discovery of Anfinsen, and various theories and models have been proposed for what has come to be known as the "*Protein Folding Problem*," our understanding of the same is far from complete.[1] The purpose of this paper is to address issues that are pertinent to the folding problem using a treatment that is analogous to that given by Marcus for electron transfer.[2]

# FORMULATION OF THE HYPOTHESIS

## Parabolic approximation

Consider the denatured state ensemble (DSE) of a spontaneously-folding fixed two-state folder at equilibrium under folding conditions wherein the variables such as temperature, pressure, pH, ionic strength etc. are defined and constant.[3,4] The solvent accessible surface area (SASA) and the Gibbs energy of each one of the conformers that comprise the DSE, and consequently, the mean SASA and Gibbs energy of the ensemble will be determined by a complex interplay of intra-protein and protein-solvent interactions (hydrogen bonds, van der Waals and electrostatic interactions, salt bridges etc.).[5-8] At finite but constant temperature, the incessant transfer of momentum from the thermal motion of water causes the polypeptide to constantly drift from its mean SASA.[9] As the chain expands, there is a favourable gain in *chain entropy* due to the increased backbone and side-chain conformational freedom, and a favourable gain in *solvation enthalpy* due to the increased solvation of the backbone and the side-chains; however, this is offset by the loss of favourable *chain enthalpy* that stems from the intra-protein backbone and the side-chain interactions, and the unfavourable decrease in *solvent entropy*, since more water molecules are now tied down by the relatively more exposed hydrophobic residues, hydrogen-bond donors and acceptors, and charged residues in the polypeptide. Conversely, as the chain attempts to become increasingly compact, there is a favourable gain in chain enthalpy due to an increase in the number of residual interactions, and a favourable increase in the solvent entropy due to the release of bound water molecules; however, this is opposed by the unfavourable decrease in both the backbone and the side-chain entropy (excluded volume entropy) and the enthalpy of desolvation.[10,11] Therefore, it is postulated that the restoring force experienced by each one of the conformers in the DSE would be proportional to their displacement from the mean SASA of the DSE along the SASA-reaction coordinate (SASA-RC), or $F(x_i) \propto (x_i - \bar{x}_{DSE})$ where $x_i$ is the SASA of the $i^{th}$ conformer in the DSE, $\bar{x}_{DSE}$ is the mean SASA of the DSE, and $F(x_i)$ is the restoring force experienced by it. Consequently, the Gibbs energy of the conformer, $G(x_i)$, is proportional to the square of this displacement, or $G(x_i) \propto (x_i - \bar{x}_{DSE})^2$. If the totality of forces that resist expansion and compaction of the polypeptide chain are assumed to be equal, then to a first approximation the conformers in the DSE may be treated as being confined to a harmonic Gibbs energy-well with a defined force constant (**Figure 1A**). Once the Gibbs energies of the conformers are known, the probabilities of their occurrence within the ensemble at equilibrium can be readily ascertained using the Boltzmann distribution law (**Figure 1B**). We will come back to this later.

The native state ensemble (NSE) in solution may be treated in an analogous manner: Although the NSE is incredibly far more structurally homogeneous than the DSE, and is sometimes treated as being equivalent to a single state (i.e., the conformational entropy of the NSE is set to zero) for the purpose of estimating the difference in conformational entropy between the DSE and the NSE, the NSE by definition is an ensemble of structures.[12-14] In fact, this thermal-noise-induced tendency to oscillate is so strong that native-folded proteins even when constrained by a crystal lattice can perform this motion.[15] Thus, at finite temperature the NSE is defined by its mean SASA ($\bar{x}_{NSE}$) and its ensemble-averaged Gibbs energy. As the native conformer attempts to become increasingly compact, its excluded volume entropy rises tremendously since most of the space in the protein core has already been occupied by the polypeptide backbone and the side-chains of the constituent amino acids.[16] In contrast, any attempt by the polypeptide chain to expand and consequently expose more SASA is met with resistance by the multitude of interactions that keep the folded structure intact. Therefore, it is postulated that the restoring force would be proportional to the displacement of the native conformer from $\bar{x}_{NSE}$ along the SASA-RC, or $F(y_i) \propto (y_i - \bar{x}_{NSE}) \Rightarrow G(y_i) \propto (y_i - \bar{x}_{NSE})^2$ where $y_i$ is the SASA of the $i^{th}$ conformer in the NSE, $F(y_i)$ is the restoring force experienced by it, and $G(y_i)$ is the Gibbs energy of the native conformer. If the sum total of the forces that resist compaction and expansion of the native conformer, respectively, are assumed to be equal in magnitude, then the conformers in the NSE may be treated as being confined to a harmonic Gibbs energy-well with a defined force constant.

## On the use of the denaturant $m_{D-N}$ value as a global reaction coordinate

The description of protein folding reactions in terms of reaction coordinates (RCs) and transition states is based on concepts borrowed from the covalent chemistry of small molecules. Because protein folding reactions are profoundly different from reactions in covalent chemistry owing to their non-covalent and multi-dimensional nature, it is often argued that their full complexity cannot be captured in sufficient detail by any single RC.[17] Nevertheless, it is not uncommon to analyse the same using one-dimensional RCs, such as the native-likeness in the backbone configuration, the fraction of native pair-wise contacts ($Q_i$) relative to the ground states DSE and NSE, the radius of gyration ($R_g$), SASA, $P_{fold}$ etc.[18,19] The use of SASA as a global RC in the proposed hypothesis poses a problem since it is very difficult, if not impossible, to accurately and precisely determine the ensemble-averaged length of the RC ($\Delta SASA_{D-N}$) using structural and/or biophysical methods. Although the mean SASA of the NSE and its fluctuations may be obtained by applying computational methods to the available crystal or solution structures of proteins,[20] such approaches are not readily applicable to the DSE.[3] Although there has been considerable progress in modelling the SASA of the DSEs using simulations,[21,22] these methods have not been used here for one predominant reason: Unlike the NSE, the residual structure in the DSEs of most proteins can be very sensitive to minor changes in the primary sequence and solvent conditions, which may not be captured effectively by these theoretical methods. Therefore, the experimentally accessible $m_{D-N}$ has been used as a proxy for the true $\Delta SASA_{D-N}$.[19]

## Postulates of the model

The Gibbs energy functions of the DSE and the NSE, denoted by $G_{DSE(r)(T)}$ and $G_{NSE(r)(T)}$ respectively, have a square-law dependence on the RC, $r$, and are described by parabolas (**Figure 2**). The curvature of parabolas is given by their respective force constants, $\alpha$ and $\omega$. As long as the primary sequence is not perturbed (*via* mutation, chemical or post-translational modification), and pressure and solvent conditions are constant, and the properties of the solvent are temperature-invariant (for example, no change in the pH due to the temperature-dependence of the p$K_a$ of the constituent buffer), the force constants $\alpha$ and $\omega$ are temperature-invariant (**Figure 3**), i.e., the conformers in the DSE and the NSE behave like linear-elastic springs. A corollary is that changes to the primary sequence, or change in solvent conditions (a change in pH, ionic strength, or addition of co-solvents) can bring about a change in either $\alpha$ or $\omega$ or both.

The vertices of the DSE and NSE-parabolas, denoted by $G_{D(T)}$ and $G_{N(T)}$, respectively, represent their ensemble-averaged Gibbs energies. Consequently, in a parabolic representation, the difference in Gibbs energy between the DSE and NSE at equilibrium is given by separation between $G_{D(T)}$ and $G_{N(T)}$ along the ordinate ($\Delta G_{D-N(T)} = G_{D(T)} - G_{N(T)}$). A decrease or an increase in $\Delta G_{D-N(T)}$ relative to the standard state/wild type upon perturbation is synonymous with the net movement of the vertices of the parabolas towards each other or away from each other, respectively, along the ordinate (**Figure 3**). Thus, a decrease in $\Delta G_{D-N(T)}$ can be due to a stabilized DSE or a destabilized NSE or both. Conversely, an increase in $\Delta G_{D-N(T)}$ can be due to a destabilized DSE or a stabilized NSE or both.

The mean length of the RC is given by the separation between $G_{D(T)}$ and $G_{N(T)}$ along the abscissa, and is identical to the experimentally accessible $m_{D-N}$ (**Figure 2C**). For the folding reaction $D \rightleftharpoons N$, since the RC increases linearly from $0 \rightarrow m_{D-N}$ in the left-to-right direction, the vertex of the DSE-parabola is always at zero along the abscissa while that of the NSE-parabola is always at $m_{D-N}$. An increase or decrease in $\Delta SASA_{D-N}$, relative to a reference state or the wild type, in accordance with the standard paradigm, will manifest as an increase or a decrease in $m_{D-N}$, respectively.[19] In a parabolic representation, an increase in $m_{D-N}$ is synonymous with the net movement of vertices of the DSE and NSE-parabolas away from each other along the abscissa. Conversely, a decrease in $m_{D-N}$ is synonymous with the net movement of the parabolas towards each other along the abscissa (**Figure 4**). As long as the primary sequence is not perturbed, and pressure and solvent conditions are constant, and the properties of the solvent are temperature-invariant, $\bar{x}_{DSE}$ and $\bar{x}_{NSE}$ are invariant with temperature, leading to $\Delta SASA_{D-N}$ being temperature-independent; consequently, the mean length of the RC, $m_{D-N}$, for a fixed two-state folder is also invariant with temperature. A corollary is that perturbations such as changes to the primary sequence *via* mutation, chemical or post-translational modification, change in pressure, pH, ionic strength, or addition of co-solvents can bring about a change in either $\bar{x}_{\text{DSE}}$, or $\bar{x}_{\text{NSE}}$, or both, leading to a change in $\Delta\text{SASA}_{\text{D-N}}$, and consequently, a change in $m_{\text{D-N}}$. Because by postulate $m_{\text{D-N}}$ is invariant with temperature, a logical extension is that for a fixed two-state folder, the ensemble-averaged difference in heat capacity between DSE and the NSE ($\Delta C_{p\text{D-N}} = C_{p\text{D}(T)} - C_{p\text{N}(T)}$) must also be temperature-invariant since these two parameters are directly proportional to each other (see discussion on the temperature-invariance of $\Delta\text{SASA}_{\text{D-N}}$, $m_{\text{D-N}}$ and $\Delta C_{p\text{D-N}}$).[23,24]

The mean position of the transition state ensemble (TSE) along the RC, $r_{\ddagger(T)}$, and the ensemble-averaged Gibbs energy of the TSE ($G_{\text{TS}(T)}$) are determined by the intersection of $G_{\text{DSE}(r)(T)}$ and $G_{\text{NSE}(r)(T)}$ functions. In a parabolic representation, the difference in SASA between the DSE and the TSE is given by the separation between $G_{\text{D}(T)}$ and the *curve-crossing* along the abscissa and is identical to $m_{\text{TS-D}(T)}$. Thus, if the mean SASA of the TSE is denoted by $\bar{x}_{\text{TSE}(T)}$, then $m_{\text{TS-D}(T)}$ is a true proxy for $\bar{x}_{\text{DSE}} - \bar{x}_{\text{TSE}(T)} = \Delta\text{SASA}_{\text{D-TS}(T)}$ and is always greater than zero no matter what the temperature. Similarly, the difference in SASA between the TSE and the NSE is given by the separation between $G_{\text{N}(T)}$ and the *curve-crossing* along the abscissa and is identical to $m_{\text{TS-N}(T)}$, i.e., $m_{\text{TS-N}(T)}$ is a true proxy for $\bar{x}_{\text{TSE}(T)} - \bar{x}_{\text{NSE}} = \Delta\text{SASA}_{\text{TS-N}(T)}$. However, unlike $m_{\text{TS-D}(T)}$ which is always greater than zero, $m_{\text{TS-N}(T)}$ can approach zero (when $\bar{x}_{\text{TSE}(T)} = \bar{x}_{\text{NSE}}$) and even become negative ($\bar{x}_{\text{TSE}(T)} < \bar{x}_{\text{NSE}}$) at very low and high temperatures for certain proteins. The ensemble-averaged Gibbs activation energy for folding is given by the separation between $G_{\text{D}(T)}$ and the *curve-crossing* along the ordinate ($\Delta G_{\text{TS-D}(T)} = G_{\text{TS}(T)} - G_{\text{D}(T)}$), and the ensemble-averaged Gibbs activation energy for unfolding is given by the separation between $G_{\text{N}(T)}$ and the *curve-crossing* along the ordinate ($\Delta G_{\text{TS-N}(T)} = G_{\text{TS}(T)} - G_{\text{N}(T)}$). The position of the *curve-crossing* along the abscissa and ordinate relative to the ground states is purely a function of the primary sequence when temperature, pressure and solvent conditions are defined. A corollary of this is that for any two-state folder, any perturbation that brings about a change in the curvature of the parabolas or the mean length of the RC can lead to a change in $m_{\text{TS-D}(T)}$. Because $m_{\text{D-N}} = m_{\text{TS-D}(T)} + m_{\text{TS-N}(T)}$ for a two-state system, any perturbation that causes an increase in $m_{\text{TS-D}(T)}$ without a change $m_{\text{D-N}}$ will concomitantly lead to a decrease in $m_{\text{TS-N}(T)}$, and *vice versa*. Consequently, the normalized solvent RCs $\beta_{T(fold)(T)} = m_{TS-D(T)}/m_{D-N}$ and $\beta_{T(unfold)(T)} = m_{TS-N(T)}/m_{D-N}$ will also vary with the said perturbation.[25]

Thus, from the postulates of the parabolic hypothesis we have three fundamentally important equations for fixed two-state protein folders:

$$\Delta G_{TS-D(T)} = \alpha \left( m_{TS-D(T)} \right)^2 \tag{1}$$

$$\Delta G_{TS-N(T)} = \omega \left( m_{TS-N(T)} \right)^2 = \omega \left( m_{D-N} - m_{TS-D(T)} \right)^2 \tag{2}$$

$$\Delta G_{D-N(T)} = \Delta G_{TS-N(T)} - \Delta G_{TS-D(T)} = \omega \left( m_{TS-N(T)} \right)^2 - \alpha \left( m_{TS-D(T)} \right)^2 \tag{3}$$

Consequently, for two-state proteins under folding conditions, as long as $\Delta G_{TS-N(T)} > \Delta G_{TS-D(T)}$ (i.e., $\Delta G_{D-N(T)} > 0$ or $\Delta G_{N-D(T)} < 0$) and $m_{TS-D(T)} > m_{TS-N(T)}$ we have the logical condition $\omega > \alpha$ (**Figure 2C**).

## Expression for the mean position of the TSE

Consider the conventional barrier-limited interconversion of the conformers in the DSE and NSE of a two-state folder at any given temperature, pressure and solvent conditions (**Figure 2C**). Because by postulate the Gibbs energy functions $G_{DSE(r)(T)}$ and $G_{NSE(r)(T)}$ have a square-law dependence on the RC, $r$, whose ensemble-averaged length is given by $m_{D-N}$, and since the RC increases linearly from $0 \to m_{D-N}$ in the left to right direction, we can write

$$G_{DSE(r)(T)} = \alpha \left( 0 - r_{(T)} \right)^2 = \alpha r_{(T)}^2 \tag{4}$$

$$G_{NSE(r)(T)} = \omega \left( m_{D-N} - r_{(T)} \right)^2 - \Delta G_{D-N(T)} \tag{5}$$

If the units of the ordinate are in kcal.mol$^{-1}$ and the RC in kcal.mol$^{-1}$.M$^{-1}$, then by definition the force constants $\alpha$ and $\omega$ have the units M$^2$.mol.kcal$^{-1}$. The mean position of the TSE along the abscissa ($r_{\ddagger(T)}$) is determined by the intersection of $G_{DSE(r)(T)}$ and $G_{NSE(r)(T)}$. Therefore, at the *curve-crossing* we have

$$G_{DSE(r\ddagger)(T)} = G_{NSE(r\ddagger)(T)} \Rightarrow \alpha r_{\ddagger(T)}^2 = \omega \left( m_{D-N} - r_{\ddagger(T)} \right)^2 - \Delta G_{D-N(T)} \tag{6}$$

$$\Rightarrow (\omega-\alpha)r_{\ddagger(T)}^2 - 2\omega m_{D-N}r_{\ddagger(T)} + \omega(m_{D-N})^2 - \Delta G_{D-N(T)} = 0 \qquad (7)$$

Solving for $r_{\ddagger(T)}$ gives (see **Appendix**)

$$r_{\ddagger(T)} \equiv m_{TS-D(T)} = \frac{\omega m_{D-N} - \sqrt{\alpha\omega(m_{D-N})^2 + \Delta G_{D-N(T)}(\omega-\alpha)}}{(\omega-\alpha)} \qquad (8)$$

$$\Rightarrow m_{TS-D(T)} = \frac{\omega m_{D-N} - \sqrt{\varphi}}{(\omega-\alpha)} \qquad (9)$$

where the discriminant $\varphi = \lambda\omega + \Delta G_{D-N(T)}(\omega-\alpha)$, and the parameter $\lambda = \alpha(m_{D-N})^2$ is analogous to the "*Marcus reorganization energy*," and by definition is the Gibbs energy required to compress the denatured polypeptide under folding conditions to a state whose SASA is identical to that of the native folded protein but without the stabilizing native interactions (**Figure 5**). Since $\alpha$ and $m_{D-N}$ are by postulate temperature-invariant, $\lambda$ is temperature-invariant by extension and depends purely on the primary sequence for a given pressure and solvent conditions. Since $m_{D-N} = m_{TS-D(T)} + m_{TS-N(T)}$ for a two-state system, we have

$$m_{TS-N(T)} = \frac{\sqrt{\varphi} - \alpha m_{D-N}}{(\omega-\alpha)} \qquad (10)$$

If the values of the force constants $\alpha$ and $\omega$, $m_{D-N}$ and $\Delta G_{D-N(T)}$ of a two-state system at any given temperature, pressure and solvent conditions are known, we can readily calculate the absolute Gibbs activation energies for the folding and unfolding (Eqs. (1) and (2)).

**Equations for the folding and the unfolding rate constants**

The two theories that feature prominently in the analyses of protein folding kinetics are the transition state theory (TST) and the Kramers' theory under high friction limit.[26-28] Despite their profound differences what is common to both is the exponential term or the Boltzmann factor. Therefore, we will start with the conventional Arrhenius expression for the rate constants (the complexity of the prefactor which here is assumed to be temperature-invariant is addressed elsewhere). Substituting Eqs. (1) and (9), and (2) and (10) in the expressions for the rate constants for folding ($k_{f(T)}$) and unfolding ($k_{u(T)}$), respectively, gives

$$k_{f(T)} = k^0 \exp\left(-\frac{\Delta G_{\text{TS-D}(T)}}{RT}\right) = k^0 \exp\left(-\frac{\alpha\left(m_{\text{TS-D}(T)}\right)^2}{RT}\right) = k^0 \exp\left(-\frac{\alpha\left(\omega m_{\text{D-N}} - \sqrt{\varphi}\right)^2}{RT(\omega-\alpha)^2}\right) \quad (11)$$

$$k_{u(T)} = k^0 \exp\left(-\frac{\Delta G_{\text{TS-N}(T)}}{RT}\right) = k^0 \exp\left(-\frac{\omega\left(m_{\text{TS-N}(T)}\right)^2}{RT}\right) = k^0 \exp\left(-\frac{\omega\left(\sqrt{\varphi} - \alpha m_{\text{D-N}}\right)^2}{RT(\omega-\alpha)^2}\right) \quad (12)$$

where $k^0$ is the pre-exponential factor with units identical to those of the rate constants (s$^{-1}$). Because the principle of microscopic reversibility stipulates that for a two-state system the ratio of the folding and unfolding rate constants must be identical to the independently measured equilibrium constant, the prefactors in Eqs. (11) and (12) must be identical.[29] Eqs. (11) and (12) may further be recast in terms of $\beta_{\text{T(fold)}(T)}$ and $\beta_{\text{T(unfold)}(T)}$ to give

$$k_{f(T)} = k^0 \exp\left(-\frac{\lambda \beta^2_{\text{T(fold)}(T)}}{RT}\right) \quad (13)$$

$$k_{u(T)} = k^0 \exp\left(-\frac{\lambda \omega \beta^2_{\text{T(unfold)}(T)}}{\alpha RT}\right) \quad (14)$$

Eqs. (11) – (14) at once demonstrate that the relationship between the rate constants, the equilibrium stability, and the denaturant *m* value is incredibly complex since the parameters in the said equations can all change depending on the nature of the perturbation and will be explored in detail elsewhere.

## The force constants are inversely proportional to the variances of the Gaussian distribution of the conformers

If $G_{\text{D}i(T)}$ and $G_{\text{N}j(T)}$ ($i, j = 1.....n$) denote the Gibbs energies of the conformers in the DSE and the NSE, respectively, then the probability distribution of their conformers along the RC at equilibrium is given by the Boltzmann law. Because by postulate the Gibbs energies of the conformers in the DSE and the NSE have a square-law dependence on the RC, *r*, whose ensemble-averaged length is given by $m_{\text{D-N}}$, and because the RC increases linearly from 0 → $m_{\text{D-N}}$ in the left-to-right direction, we can write

$$p_{Di(T)} = \frac{1}{Q_{DSE(T)}} \exp\left(-\frac{G_{Di(T)}}{RT}\right) = \frac{1}{Q_{DSE(T)}} \exp\left(-\frac{\alpha r_{(T)}^2}{RT}\right) \quad (15)$$

$$p_{Nj(T)} = \frac{1}{Q_{NSE(T)}} \exp\left(-\frac{G_{Nj(T)}}{RT}\right) = \frac{1}{Q_{NSE(T)}} \exp\left(-\frac{\omega(m_{D-N} - r_{(T)})^2 - \Delta G_{D-N(T)}}{RT}\right) \quad (16)$$

where $p_{Di(T)}$ and $p_{Nj(T)}$ denote the Boltzmann probabilities of the conformers in the DSE and the NSE, respectively, with their corresponding partition functions $Q_{DSE(T)}$ and $Q_{NSE(T)}$ being given by

$$Q_{DSE(T)} = \sum_{i=1}^{n} \exp\left(-\frac{G_{Di(T)}}{RT}\right) = \int_{-\infty}^{\infty} \exp\left(-\frac{\alpha r_{(T)}^2}{RT}\right) dr = \sqrt{\frac{\pi RT}{\alpha}} \quad (17)$$

$$Q_{NSE(T)} = \sum_{j=1}^{n} \exp\left(-\frac{G_{Nj(T)}}{RT}\right) = \int_{-\infty}^{\infty} \exp\left(\frac{\Delta G_{D-N(T)} - \omega(m_{D-N} - r_{(T)})^2}{RT}\right) dr$$
$$= \sqrt{\frac{\pi RT}{\omega}} \exp\left(\frac{\Delta G_{D-N(T)}}{RT}\right) \quad (18)$$

Because the equilibrium is dynamic, there is always a constant thermal noise-driven flux of the conformers from the DSE to the NSE, and from the NSE to the DSE, *via* the TSE. Consequently, there is always a constant albeit incredibly small population of conformers in the TSE at equilibrium. Now consider the first-half of a protein folding reaction as shown in Scheme 1, where [*D*], [*TS*] and [*N*] denote the equilibrium concentrations of the DSE, the TSE, and the NSE, respectively, in molar.

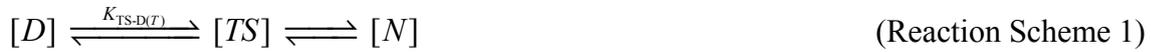

$[D] \underset{}{\overset{K_{TS-D(T)}}{\rightleftharpoons}} [TS] \rightleftharpoons [N]$  (Reaction Scheme 1)

From the perspective of a folding reaction, the conformers in the activated state or the TSE may be thought of as a subset of denatured conformers with very high Gibbs energies. Therefore, we may assume that the conformers in the TSE are in equilibrium with those conformers that are at the bottom of the denatured Gibbs basin. If $G_{D(T)}$, $G_{N(T)}$ and $G_{TS(T)}$ denote the mean Gibbs energies of the DSE, NSE, and the TSE, respectively, then the ratio of the molar concentration of the conformers at the bottom of the denatured Gibbs basin and those in the TSE is given by

$$-RT\ln\left(\frac{[TS]}{[D]}\right) = \Delta G_{\text{TS-D}(T)} \Rightarrow \frac{[TS]}{[D]} = \exp\left(-\frac{\Delta G_{\text{TS-D}(T)}}{RT}\right) = \exp\left(-\frac{\alpha\left(m_{\text{TS-D}(T)}\right)^2}{RT}\right) \quad (19)$$

Similarly for the partial unfolding reaction (Reaction Scheme 2), the conformers in the TSE may be thought of as a subset of native conformers with very high Gibbs energies. Therefore, we may write

$$[N] \xrightleftharpoons{K_{\text{TS-N}(T)}} [TS] \rightleftharpoons [D] \qquad \text{(Reaction Scheme 2)}$$

$$-RT\ln\left(\frac{[TS]}{[N]}\right) = \Delta G_{\text{TS-N}(T)} \Rightarrow \frac{[TS]}{[N]} = \exp\left(-\frac{\Delta G_{\text{TS-N}(T)}}{RT}\right) = \exp\left(-\frac{\omega\left(m_{\text{TS-N}(T)}\right)^2}{RT}\right) \quad (20)$$

Because the SASA of the conformers in the DSE or the NSE is determined by a multitude of intra-protein and protein-solvent interactions, we may invoke the central limit theorem and assume that the distribution of the SASA of the conformers is a Gaussian. If $\sigma^2_{\text{DSE}(T)}$ and $\sigma^2_{\text{NSE}(T)}$ denote the variances of the DSE and the NSE-Gaussian probability density functions (Gaussian-PDFs), respectively, along the SASA-RC which in our case is its proxy, the experimentally measurable and temperature-invariant $m_{\text{D-N}}$, and $\bar{x}_{\text{DSE}}$, $\bar{x}_{\text{NSE}}$, and $\bar{x}_{\text{TSE}(T)}$ denote the mean SASAs of the DSE, the NSE, and the TSE, respectively, then the ratio of the molar concentration of the conformers whose SASA is identical to the mean SASA of the DSE to those whose SASA is identical to the mean SASA of the TSE is given by

$$\frac{[TS]}{[D]} = \frac{\sqrt{2\pi\sigma^2_{\text{DSE}(T)}}}{\sqrt{2\pi\sigma^2_{\text{DSE}(T)}}}\exp\left(-\frac{\left(\bar{x}_{\text{TSE}(T)} - \bar{x}_{\text{DSE}}\right)^2}{2\sigma^2_{\text{DSE}(T)}}\right) = \exp\left(-\frac{\left(m_{\text{TS-D}(T)}\right)^2}{2\sigma^2_{\text{DSE}(T)}}\right) \quad (21)$$

Similarly, the ratio of the molar concentration of the conformers whose SASA is identical to the mean SASA of the TSE to those whose SASA is identical to the mean SASA of the NSE is given by

$$\frac{[TS]}{[N]} = \frac{\sqrt{2\pi\sigma^2_{\text{NSE}(T)}}}{\sqrt{2\pi\sigma^2_{\text{NSE}(T)}}}\exp\left(-\frac{\left(\bar{x}_{\text{TSE}(T)} - \bar{x}_{\text{NSE}}\right)^2}{2\sigma^2_{\text{NSE}(T)}}\right) = \exp\left(-\frac{\left(m_{\text{TS-N}(T)}\right)^2}{2\sigma^2_{\text{NSE}(T)}}\right) \quad (22)$$

Because the ratio of the conformers in the TSE to those in the ground states must be the same whether we use a Gaussian approximation or the Boltzmann distribution (compare Eqs. (19) and (21), and Eqs. (20) and (22)), we can write

$$\exp\left(-\frac{\alpha\left(m_{\text{TS-D}(T)}\right)^2}{RT}\right) = \exp\left(-\frac{\left(m_{\text{TS-D}(T)}\right)^2}{2\sigma^2_{\text{DSE}(T)}}\right) \Rightarrow \sigma^2_{\text{DSE}(T)} = \frac{RT}{2\alpha} \tag{23}$$

$$\exp\left(-\frac{\omega\left(m_{\text{TS-N}(T)}\right)^2}{RT}\right) = \exp\left(-\frac{\left(m_{\text{TS-N}(T)}\right)^2}{2\sigma^2_{\text{NSE}(T)}}\right) \Rightarrow \sigma^2_{\text{NSE}(T)} = \frac{RT}{2\omega} \tag{24}$$

Thus, for any two state folder at constant temperature, pressure, and solvent conditions, the variance of the Gaussian distribution of the conformers in the DSE or the NSE along the $m_{\text{D-N}}$ RC is inversely proportional to their respective force constants; and for a two-state system with given force constants, the variance is directly proportional to the absolute temperature. Naturally, in the absence of thermal energy ($T = 0$ K), all classical motion will cease and $\sigma^2_{\text{DSE}(T)} = \sigma^2_{\text{NSE}(T)} = 0$. The relationship between protein motion and function will be explored elsewhere. The area enclosed by the DSE and the NSE-Gaussians is given by

$$I_{\text{DSE}(T)} = \int_{-\infty}^{\infty} \exp(-ax^2)\,dx = \sqrt{\frac{\pi}{a}} = \sqrt{2\pi\sigma^2_{\text{DSE}(T)}} = \sqrt{\frac{\pi RT}{\alpha}} = Q_{\text{DSE}(T)} \tag{25}$$

$$I_{\text{NSE}(T)} = \int_{-\infty}^{\infty} \exp(-by^2)\,dy = \sqrt{\frac{\pi}{b}} = \sqrt{2\pi\sigma^2_{\text{NSE}(T)}} = \sqrt{\frac{\pi RT}{\omega}} = Q_{\text{NSE}(T)} \exp\left(-\frac{\Delta G_{\text{D-N}(T)}}{RT}\right) \tag{26}$$

where $x = (x_i - \bar{x}_{\text{DSE}})$, $y = (y_i - \bar{x}_{\text{NSE}})$, $x_i$ and $y_i$ denote the SASAs of the $i^{\text{th}}$ conformers in the DSE and the NSE, respectively, $a = 1/2\sigma^2_{\text{DSE}(T)}$, $b = 1/2\sigma^2_{\text{NSE}(T)}$, and $I_{\text{DSE}(T)}$ and $I_{\text{NSE}(T)}$ denote the areas enclosed by the DSE and NSE-Gaussians, respectively, along the $m_{\text{D-N}}$ RC. The reader will note that for a polypeptide of finite length, the maximum permissible SASA is determined by the fully extended chain and the minimum by the excluded volume entropy. Thus, the use of the limits $-\infty$ to $+\infty$ in Eqs. (17), (18), (25) and (26) is not physically justified. However, because the populations decrease exponentially as the conformers in both the DSE and the NSE are displaced from their mean SASA, the difference in the magnitude of the partition functions calculated using actual limits versus $-\infty$ and $+\infty$ will be insignificant. Eqs. (23) and (24) allow $k_{f(T)}$ and $k_{u(T)}$ to be recast in terms of the variances of the DSE and the NSE-Gaussians

$$k_{f(T)} = k^0 \exp\left(-\frac{(m_{\text{TS-D}(T)})^2}{2\sigma^2_{\text{DSE}(T)}}\right) \tag{27}$$

$$k_{u(T)} = k^0 \exp\left(-\frac{(m_{\text{TS-N}(T)})^2}{2\sigma^2_{\text{NSE}(T)}}\right) \tag{28}$$

We will show elsewhere when we deal with non-Arrhenius kinetics in protein folding in detail that although the variance of the DSE and the NSE-Gaussians increases linearly with absolute temperature, the *curve-crossing* and the Gibbs barrier heights for folding and unfolding are non-linear functions of their respective variances.

**Equations for equilibrium stability**

The relationship between the partition functions, the area enclosed by DSE and the NSE Gaussians, and the Gibbs energy of unfolding may be readily obtained by dividing Eq. (18) by (17)

$$\Delta G_{\text{D-N}(T)} = RT \ln\left(\sqrt{\frac{\omega}{\alpha}} \frac{Q_{\text{NSE}(T)}}{Q_{\text{DSE}(T)}}\right) = RT \ln\left(\frac{\sigma_{\text{DSE}(T)} Q_{\text{NSE}(T)}}{\sigma_{\text{NSE}(T)} Q_{\text{DSE}(T)}}\right) = RT \ln\left(\frac{Q_{\text{NSE}(T)}}{I_{\text{NSE}(T)}}\right) \tag{29}$$

$$\therefore \frac{I_{\text{DSE}(T)}}{I_{\text{NSE}(T)}} = \sqrt{\frac{2\pi\sigma^2_{\text{DSE}(T)}}{2\pi\sigma^2_{\text{NSE}(T)}}} = \frac{\sigma_{\text{DSE}(T)}}{\sigma_{\text{NSE}(T)}} = \sqrt{\frac{\omega}{\alpha}} \tag{30}$$

where $\sigma_{\text{DSE}(T)}$ and $\sigma_{\text{NSE}(T)}$ denote the standard deviations of the DSE and NSE-Gaussians, respectively, along the $m_{\text{D-N}}$ RC. There are many other ways of recasting the equation for equilibrium stability (not shown), but the simplest and perhaps the most useful form is

$$\Delta G_{\text{D-N}(T)} = \lambda\left(\frac{\omega}{\alpha}\beta^2_{\text{T(unfold)}(T)} - \beta^2_{\text{T(fold)}(T)}\right) = \lambda\left(\frac{\sigma^2_{\text{DSE}(T)}}{\sigma^2_{\text{NSE}(T)}}\beta^2_{\text{T(unfold)}(T)} - \beta^2_{\text{T(fold)}(T)}\right) \tag{31}$$

Eq. (31) demonstrates that when pressure, temperature, and solvent conditions are constant, the equilibrium and kinetic behaviour of those proteins that fold spontaneously without the need for any accessory factors is determined purely by three primary-sequence-dependent variables which are: (*i*) the ensemble-averaged mean and variance of the Gaussian distribution of the conformers in the DSE along SASA-reaction-coordinate; (*ii*) the ensemble-averaged mean and variance of the Gaussian distribution of the conformers in the NSE along the SASA-reaction-coordinate; and (*iii*) the position of the *curve-crossing* along the abscissa. A necessary consequence of Eq. (31) is that: (*i*) if for spontaneously-folding fixed two-state systems at constant pressure and solvent conditions $\Delta SASA_{D-N}$ is positive and temperature-invariant (i.e., $m_{D-N}$ and $\Delta C_{pD-N}$ are temperature-invariant), and $\beta_{T(fold)(T)} \geq 0.5$ when $T = T_S$ (the temperature at which stability is a maximum),[30] then it is impossible for such systems to be stable at equilibrium ($\Delta G_{D-N(T)} > 0$) unless $\sigma^2_{DSE(T)} > \sigma^2_{NSE(T)}$ no matter what the temperature; (*ii*) if two related or unrelated two-state systems have identical pair of force constants, and if their $\Delta SASA_{D-N}$ as well as the absolute position of the DSE and the NSE along the SASA-RC are also identical, then the protein which folds through a more solvent-exposed TSE will be more stable at equilibrium; and (*iii*) if $m_{D-N}$ and $\Delta C_{pD-N}$ are temperature-invariant, a spontaneously-folding two-state system at constant pressure and solvent conditions, irrespective of its primary sequence or 3-dimensional structure, will be maximally stable at equilibrium when its denatured conformers are displaced the least from the mean of their ensemble to reach the TSE along the SASA-RC (the *principle of least displacement*). Because equilibrium stability is the greatest at $T_S$, a logical extension is that $m_{TS-D(T)}$ or $\beta_{T(fold)(T)}$ must be a minimum, and $m_{TS-N(T)}$ or $\beta_{T(unfold)(T)}$ a maximum at $T_S$ (**Figure 3**). A corollary is that the Gibbs activation barriers for folding and unfolding are a minimum and a maximum, respectively, when the difference in SASA between the DSE and the TSE is the least. Mathematical formalism for why the activation entropies for folding and unfolding must both be zero at $T_S$ will be shown in the subsequent publication.

The correspondence between Gibbs parabolas and Gaussian-PDFs for two well-studied two-state proteins: (*i*) CI2; and (*ii*) the B domain of staphylococcal protein A (BdpA Y15W) are shown in **Figures 6** and **7**, respectively. The parameters required to generate these figures are given in their legends. As mentioned earlier, the logical condition that as long as $\Delta G_{D-N(T)} > 0$ and $m_{TS-D(T)} > m_{TS-N(T)}$ then $\omega > \alpha$ is readily apparent from **Figures 6A** and **7A**. Because the Gaussian variances of the DSE and the NSE are inversely proportional to the force constants, $\omega > \alpha$ implies $\sigma^2_{NSE(T)} < \sigma^2_{DSE(T)}$ (**Figures 6B** and **7B**). A detailed discussion of the theory underlying the procedure required to extract the values of the force constants from the chevrons and its inherent limitations is beyond the scope of this article since it involves a radical reinterpretation of the chevron. A brief description is given in methods.

## On the temperature-invariance of $\Delta SASA_{D-N}$, $m_{D-N}$ and $\Delta C_{pD-N}$

One of the defining postulates of the parabolic hypothesis is that for a *spontaneously-folding fixed two-state folder*, as long as the primary sequence is not perturbed *via* mutation, chemical or post-translational modification, and pressure and solvent conditions are constant, and the properties of the solvent are invariant with temperature, the ensemble-averaged SASAs of the DSE and NSE, to a first approximation, are temperature-invariant; consequently, the dependent variables $m_{D-N}$ and $\Delta C_{pD-N}$ will also be temperature-invariant.

Consider the DSE of a two-state folder at equilibrium under folding conditions: Within the steric and energetic constraints imposed by intra-chain and chain-solvent interactions, the SASA of the denatured conformers will be normally distributed with a defined mean ($\bar{x}_{DSE}$) and variance ($\sigma^2_{DSE(T)}$). Now, if we raise the temperature of the system by tiny amount $\delta T$ such that the new temperature is $T+\delta T$, a tiny fraction of the conformers will be displaced from the mean of the ensemble, some with SASA that is greater than the mean, and some with SASA that is less than the mean; and the magnitude of this displacement from the ensemble-mean will be determined by the force constant. Consequently, there will be a tiny increase in the variance of the Gaussian distribution, and a new equilibrium will be established. Thus, as long as the integrity of the spring (i.e., the primary sequence) is not compromised, and pressure and solvent conditions are constant, the distribution itself will not be biased in any one particular way or another, i.e., the number of conformers that have become more expanded than the mean of the ensemble, on average, will be identical to the number of conformers that have become more compact than the ensemble-mean, leading to $\bar{x}_{DSE}$ being invariant with temperature. A similar argument may be applied to the NSE leading to the conclusion that although its variance increases linearly with temperature, its mean SASA ($\bar{x}_{NSE}$) will be temperature-invariant. However, if the molecular forces that resist expansion and compaction of the conformers in the DSE are not equal or approximately equal and change with temperature, then the assumption that the conformers in the DSE are confined to a harmonic Gibbs energy-well would be flawed. What is implied by this is the distribution of the conformers in the DSE along the SASA-RC is no longer a Gaussian, but instead a skewed Gaussian. For example, if the change in temperature causes a shift in the balance of molecular forces such that it is relatively easier for the denatured conformer to expand rather than become compact, in a parabolic representation, the left arm of the DSE-parabola will be shallow as compared to the right arm, and the Gaussian distribution will be negatively skewed, leading to a shift in the mean of the distribution to the left. In other words, $\bar{x}_{DSE}$ will increase, and assuming that $\bar{x}_{NSE}$ is temperature-invariant, will lead to an increase in $\Delta SASA_{D-N}$, and by extension, an increase in $m_{D-N}$. In contrast, if the change in temperature makes it easier for the denatured conformer to become compact rather than expand, then the right arm of the DSE-parabola will become shallow as compared to the left arm, and the Gaussian distribution will become positively skewed; consequently, $\bar{x}_{DSE}$ will decrease leading to a decrease in $\Delta SASA_{D-N}$ and $m_{D-N}$. Similar arguments apply to the NSE. Thus, as long as the Gibbs energy-wells are harmonic and their force constants are temperature-invariant, $\Delta SASA_{D-N}$, $m_{D-N}$, and $\Delta C_{pD-N}$ will be temperature-invariant.

The approximation that the mean length of the RC is invariant with temperature is supported by both theory and experiment: (*i*) The $R_g$ of the DSE and the NSE (after 200 pico seconds of simulation) of the truncated CI2 generated from all atom molecular dynamic simulations (MD simulations) varies little between 300 – 350 K(see Table 1 and explanation in page 214 in Lazaridis and Karplus, 1999);[31] (*ii*) Studies on the thermal expansion of native metmyoglobin by Petsko and colleagues demonstrate that the increase in the SASA and the volume of the folded protein on heating from 80 – 300 K is not more than 2 – 3%;[32] (*iii*) In chemical denaturation experiments as a function of temperature, $m_{D-N}$ is, in general, temperature-invariant within experimental error.[33-36] In addition, it is logically inconsistent to argue about possible temperature-induced changes in $m_{D-N}$ when its counterpart, $\Delta C_{pD-N}$, is assumed to be temperature-invariant in the analyses of thermal denaturation data.[30]

The widely accepted explanation for the large and positive $\Delta C_{pD-N}$ of proteins is based on Kauzmann's "*liquid-liquid transfer*" model (LLTM) which likens the hydrophobic core of the native folded protein to a liquid alkane, and the greater heat capacity of the DSE as compared to the NSE is attributed primarily to the anomalously high heat capacity and low entropy of the "clathrates" or "microscopic icebergs" of water that form around the exposed non-polar residues in the DSE (see Baldwin, 2014, and references therein).[37,38] Because the size of the solvation shell depends on the SASA of the non-polar solute, it naturally follows that the change in the heat capacity must be proportional to the change in the non-polar

SASA that accompanies a reaction. Consequently, protein unfolding reactions which are accompanied by large changes in non-polar SASA, also lead to large and positive changes in the heat capacity.[39,40] Because the denaturant $m$ values are also directly proportional to the change in SASA that accompany protein (un)folding reactions, the expectation is that $m_{D-N}$ and $\Delta C_{pD-N}$ values must also be proportional to each other: The greater the $m_{D-N}$ value, the greater is the $\Delta C_{pD-N}$ value and *vice versa*.[23,24] However, since the residual structure in the DSEs of proteins under folding conditions is both sequence and solvent-dependent (i.e., the SASAs of the DSEs two proteins of identical chain lengths but dissimilar primary sequences need not necessarily be the same even under identical solvent conditions),[3,4] and because we do not yet have reliable theoretical or experimental methods to accurately and precisely quantify the SASA of the DSEs of proteins under folding conditions (the values are model-dependent),[21,22] the data scatter in plots that show correlation between the experimentally determined $m_{D-N}$ or $\Delta C_{pD-N}$ values (which reflect the true $\Delta SASA_{D-N}$) and the theoretical model-dependent values of $\Delta SASA_{D-N}$ can be significant (see Fig. 2 in Myers et al., 1995, and Fig. 3 in Robertson and Murphy, 1997). Now, since the solvation shell around the DSEs of large proteins is relatively greater than that of small proteins even when the residual structure in the DSEs under folding conditions is taken into consideration, large proteins on average expose relatively greater amount of non-polar SASA upon unfolding than do small proteins; consequently, both $m_{D-N}$ and $\Delta C_{pD-N}$ values also correlate linearly with chain-length, albeit with considerable scatter since chain length, owing to the residual structure in the DSEs, is unlikely to be a true descriptor of the SASA of the DSEs of proteins under folding conditions (note that the scatter can also be due to certain proteins having anomalously high or low number of non-polar residues). The point we are trying to make is the following: Because the native structures of proteins are relatively insensitive to small variations in pH and co-solvents,[41-43] and since the number of ways in which foldable polypeptides can be packed into their native structures is relatively limited (as inferred from the limited number of protein folds, see SCOP: www.mrc-lmb.cam.ac.uk and CATH: www.cathdb.info databases), one might find a reasonably good correlation between chain lengths and the SASAs of the NSEs for a large dataset of proteins of differing primary sequences under varying solvents (see Fig. 1 in Miller et al., 1987).[16,44] However, since the SASAs of the DSEs under folding conditions, owing to residual structure are variable, until and unless we find a way to accurately simulate the DSEs of proteins, and if and only if these theoretical methods are sensitive to point mutations, changes in pH, co-solvents, neutral crowding agents, temperature and pressure, it is almost impossible to arrive at a universal equation that will describe how the $\Delta SASA_{D-N}$ under folding conditions will vary with chain length, and by logical extension, how $m_{D-N}$ and $\Delta C_{pD-N}$ will vary with SASA or chain length. Analyses of $\Delta C_{pD-N}$ values for a large dataset of proteins show that they generally vary between 10-20 cal.mol$^{-1}$.K$^{-1}$.residue$^{-1}$.[23,24]

Now that we have summarised the inter-relationships between $\Delta SASA_{D-N}$, $m_{D-N}$, and $\Delta C_{pD-N}$, it is easy to see that when $\Delta SASA_{D-N}$ is temperature-invariant, so too must $\Delta C_{pD-N}$, i.e., the absolute heat capacities of the DSE and the NSE may vary with temperature, but their difference, to a first approximation, can be assumed to be temperature-invariant. The reasons for this approximation are as follows: (*i*) the variation in $\Delta C_{pD-N(T)}$ over a substantial temperature range is comparable to experimental noise;[39] and (*ii*) the variation in equilibrium stability that stems from small variation in $\Delta C_{pD-N(T)}$ is once again comparable to experimental noise.[30] Consequently, the use of modified Gibbs-Helmholtz relationships with a temperature-invariant $\Delta C_{pD-N}$ term is a common practice in the field of protein folding, and is used to ascertain the temperature-dependence of the enthalpies, the entropies, and the Gibbs energies of unfolding/folding at equilibrium (Eqs. (32) – (34)). However, what is not justified is the use of experimentally determined $\Delta C_{pD-N}$ of the "wild type/reference protein" for all its mutants for the purpose of calculating the change in enthalpies, entropies and the Gibbs energies of unfolding upon mutation (i.e., $\Delta\Delta H_{D-N(wt-mut)(T)}$, $\Delta\Delta S_{D-N(wt-mut)(T)}$ and $\Delta\Delta G_{D-N(wt-mut)(T)}$; the subscripts 'wt' and 'mut' denote the wild type and the mutant protein, respectively). This is especially true if the $m_{D-N}$ values of the mutants are significantly different from that of the wild type, since those mutants with increased $m_{D-N}$ values will be expected to have increased $\Delta C_{pD-N}$ values, and *vice versa*, for identical solvent conditions and pressure, as compared to the wild type or the reference protein. These considerations are implicit in the Schellman approximation: $\Delta\Delta G_{D-N(wt-mut)(T_m)} \approx \Delta H_{D-N(wt)(T_m)} \Delta T_{m(wt-mut)} / T_{m(wt)}$ (see Fig. 8 in Becktel and Schellman, 1987, and discussion therein).

$$\Delta H_{D-N(T)} = \Delta H_{D-N(T_m)} + \int_{T_m}^{T} \Delta C_{pD-N(T)}\, dT = \Delta H_{D-N(T_m)} + \Delta C_{pD-N}\left(T - T_m\right) \qquad (32)$$

$$\Delta S_{\text{D-N}(T)} = \Delta S_{\text{D-N}(T_m)} + \int_{T_m}^{T} \frac{\Delta C_{p\text{D-N}(T)}}{T} dT = \Delta S_{\text{D-N}(T_m)} + \Delta C_{p\text{D-N}} \ln\left(\frac{T}{T_m}\right)$$
$$= \left(\frac{\Delta H_{\text{D-N}(T_m)}}{T_m}\right) + \Delta C_{p\text{D-N}} \ln\left(\frac{T}{T_m}\right) \tag{33}$$

$$\Delta G_{\text{D-N}(T)} = \Delta H_{\text{D-N}(T_m)}\left(1 - \frac{T}{T_m}\right) + \Delta C_{p\text{D-N}}(T - T_m) + T\Delta C_{p\text{D-N}} \ln\left(\frac{T_m}{T}\right) \tag{34}$$

where $\Delta H_{\text{D-N}(T)}$, $\Delta H_{\text{D-N}(T_m)}$ and $\Delta S_{\text{D-N}(T)}$, $\Delta S_{\text{D-N}(T_m)}$ denote the equilibrium enthalpies and the entropies of unfolding, respectively, at any given temperature, and at the midpoint of thermal denaturation ($T_m$), respectively, for a given two-state folder under defined solvent conditions. The temperature-invariant and the temperature-dependent difference in heat capacity between the DSE and NSE is denoted by $\Delta C_{p\text{D-N}}$ and $\Delta C_{p\text{D-N}(T)}$, respectively.

## TESTS OF HYPOTHESIS

A logical way of testing hypotheses in empirical sciences is to make quantitative predictions and verify them *via* experiment.[45] The greater the number of predictions, and the more risky they are, the more testable is the hypothesis and *vice versa*; and the greater is the agreement between theoretical prediction and experiment in such tests of hypothesis, the more certain are we of its veracity. Naturally, any hypothesis that insulates itself from "falsifiability, or refutability, or testability," is either pseudoscience or pathological science.[45,46] The theory described here readily lends itself to falsifiability because it makes certain quantitative predictions which can be immediately verified *via* experiment.

### The variation in $m_{\text{TS-D}(T)}$ and $m_{\text{TS-N}(T)}$ with $m_{\text{D-N}}$

A general observation in two-state protein folding is that whenever mutations or a change in solvent conditions cause statistically significant changes in the $m_{\text{D-N}}$ value, a large fraction or almost all of this change is manifest as a variation in $m_{\text{TS-D}(T)}$, with little or almost no change in $m_{\text{TS-N}(T)}$ (see Figs. 7 and 9 in Sanchez and Kiefhaber, 2003).[19] Although these effects were analysed using self-interaction and cross-interaction parameters,[19] the question is "Why must perturbation-induced changes in $m_{\text{D-N}}$ predominantly manifest as changes in $m_{\text{TS-D}(T)}$?" Is there any theoretical basis for this empirical observation? Importantly, can we predict how $m_{\text{TS-D}(T)}$ varies as a function of $m_{\text{D-N}}$ for any given two-state folder of a given equilibrium stability when temperature, pressure and solvent are constant? To simulate the behaviour two hypothetical two-state systems one with force constants $\alpha = 1$ and $\omega = 10$ $M^2.mol.kcal^{-1}$ (**Figure 8A**), and the other with $\alpha = 1$ and $\omega = 100$ $M^2.mol.kcal^{-1}$ (**Figure 8B**) were chosen. Within each one of these pair of parent two-state systems are six sub-systems with the same pair of force constants as the parent system but with a unique and constant $\Delta G_{D-N(T)}$. We now ask how the *curve-crossings* for each of these systems change when the separation between the vertices of DSE and NSE-parabolas along their abscissae are allowed to vary (i.e., a change in $m_{D-N}$ as in **Figure 4**). Simply put, what we are doing is taking a pair of intersecting parabolas of differing curvature such that $\omega > \alpha$ and systematically varying the separation between their vertices along the abscissa and ordinate, and calculating the position of the *curve-crossing* along the abscissa for each case according to Eqs. (9) and (10). Despite the model being very simplistic (because the curvature of the parabolas can change with structural or solvent perturbation), the simulated behaviour is strikingly similar to that of 1064 proteins from 31 two-state systems: A perturbation-induced change in $m_{D-N}$ is predominantly manifest as a change in $m_{TS-D(T)}$ with little or no change in $m_{TS-N(T)}$ (**Figure 9** and **Figure 9−figure supplement 1**). Although the apparent position of the TSE along the RC as measured by $\beta_{T(fold)(T)}$ changes, the absolute position of the TSE along the RC may not change significantly, and this effect can be particularly pronounced for systems with high $\beta_{T(fold)(T)}$ or late TSEs (**Figure 8B**). This ability to simulate the behaviour of real systems serves as the first test of the hypothesis.

## Non-Arrhenius kinetics

Unlike the temperature-dependence of the rate coefficients of most chemical reactions of small molecules, protein folding reactions are characterised by non-Arrhenius kinetics, i.e., at constant pressure and solvent conditions, $k_{f(T)}$ initially increases with an increase in temperature and reaches a plateau; and any further increase in temperature beyond this point causes $k_{f(T)}$ to decrease. This anomalous non-linear temperature-dependence of $k_{f(T)}$ has been observed in both experiment and computer simulations.[17,36,47-56] Two predominant explanations have been given for this behaviour: (*i*) non-linear temperature-dependence of the prefactor on rugged energy landscapes;[17] and (*ii*) the heat capacities of activation, $\Delta C_{pD-TS(T)}$ and $\Delta C_{pTS-N(T)}$, which in turn lead to temperature-dependent enthalpies and entropies of activation for folding and unfolding.[36,48,50,51] Arguably one of the most important and experimentally verifiable predictions of the parabolic hypothesis is that *"as long as the enthalpies and the entropies of unfolding/folding at equilibrium display a large variation with temperature, and as a consequence, equilibrium stability is a non-linear function of temperature, both $k_{f(T)}$ and $k_{u(T)}$ will have a non-linear dependence on temperature."* The equations that describe the temperature-dependence of $k_{f(T)}$ and $k_{u(T)}$ of two-state systems under constant pressure and solvent conditions may be readily derived by substituting Eq. (34) in (11) and (12).

$$k_{f(T)} = k^0 \, e^{-\frac{\alpha\left(\omega m_{D-N} - \sqrt{\lambda\omega + \left(\left(\Delta H_{D-N(T_m)}\left(1-\frac{T}{T_m}\right) + \Delta C_{pD-N}(T-T_m) + T\Delta C_{pD-N}\ln\left(\frac{T_m}{T}\right)\right)(\omega-\alpha)\right)}\right)^2}{RT(\omega-\alpha)^2}} \qquad (35)$$

$$k_{u(T)} = k^0 \, e^{-\frac{\omega\left(\sqrt{\lambda\omega + \left(\left(\Delta H_{D-N(T_m)}\left(1-\frac{T}{T_m}\right) + \Delta C_{pD-N}(T-T_m) + T\Delta C_{pD-N}\ln\left(\frac{T_m}{T}\right)\right)(\omega-\alpha)\right)} - \alpha m_{D-N}\right)^2}{RT(\omega-\alpha)^2}} \qquad (36)$$

Thus, if the parameters $\Delta H_{D-N(T_m)}$, $T_m$, $\Delta C_{pD-N}$, $m_{D-N}$, the force constants $\alpha$ and $\omega$, and $k^0$ (assumed to be temperature-invariant) are known for any given two-state system, the temperature-dependence of $k_{f(T)}$ and $k_{u(T)}$ may be readily ascertained. Why does the prediction of non-Arrhenius kinetics constitute a rigorous test of the parabolic hypothesis (see confirming evidence, Popper, 1953)? As is readily apparent, the values of the constants and variables in Eqs. (35) and (36) come from two different sources: While the values of $\alpha$ and $\omega$, $k^0$, and $m_{D-N}$ are extracted from the chevron, i.e., from the variation in $k_{f(T)}$ and $k_{u(T)}$ with denaturant at constant temperature, pressure and solvent conditions (i.e., all solvent variables excluding the denaturant are constant), $\Delta H_{D-N(T_m)}$ and $T_m$ are determined from thermal denaturation at constant pressure and identical buffer conditions as above but without the denaturant, using either calorimetry or van't Hoff analysis of a sigmoidal thermal denaturation curve, obtained for instance by monitoring the change in a suitable spectroscopic signal with temperature (typically CD 217 nm for β-sheet proteins, CD 222 nm for α-helical proteins or CD 280 nm to monitor tertiary structure).[30,57] The final parameter, $\Delta C_{pD-N}$, is once again determined independently (i.e., the slope of a plot of model-independent calorimetric $\Delta H_{D-N(T_m)(cal)}$ versus $T_m$, see Fig. 4 in Privalov, 1989). What this essentially implies is that if Eqs. (35) and (36) predict a non-linear temperature-dependence of $k_{f(T)}$ and $k_{u(T)}$, and importantly, if their absolute values agree reasonably well with experimental data, then the success of such a prediction cannot be fortuitous since it is statistically improbable for these parameters obtained from fundamentally different kinds of experiments to collude and yield the right values. We are then left with the alternative that at least to a first approximation, the hypothesis is valid.

The predictions of Eqs. (35) and (36) are shown for three well-studied two-state folders: (*i*) BdpA Y15W, the 60-residue three-helix B domain of Staphylococcal protein A (**Figure 10** and its figure supplements);[58] (*ii*) BBL H142W, the 47-residue all-helical member of the Peripheral-subunit-binding-domain family (**Figure 11** and its figure supplements);[59] and (*iii*) FBP28 WW, the 37-residue Formin-binding three-stranded β-sheet WW domain (**Figure 12** and its figure supplements).[60] Inspection of **Figures 10A, 11A** and **12A** (see also **Figure supplement 2A** for each of these figures) shows that Eq. (35) makes a remarkable prediction that $k_{f(T)}$ has a non-linear dependence on temperature. Starting from a low temperature, $k_{f(T)}$ initially increases with an increase in temperature and reaches a maximal value at $T = T_{H(\text{TS-D})}$ where $\partial \ln k_{f(T)} / \partial T = \Delta H_{\text{TS-D}(T)} / RT^2 = 0 \Rightarrow \Delta H_{\text{TS-D}(T)} = 0$; and any further increase in temperature beyond this point will cause a decrease in $k_{f(T)}$. The reader will note that the partial derivatives are purely to indicate that these relationships hold if and only if the pressure, solvent variables and the prefactor are constant.

In contrast, inspection of **Figures 10B, 11B** and **12B** (see also **Figure supplement 2B** for each of these figures) shows that $k_{u(T)}$ starting from a low temperature, decreases with a rise in temperature and reaches a minimum at $T = T_{H(\text{TS-N})}$ where $\partial \ln k_{u(T)} / \partial T = \Delta H_{\text{TS-N}(T)} / RT^2 = 0 \Rightarrow \Delta H_{\text{TS-N}(T)} = 0$; and any further increase in temperature beyond this point will cause an increase in $k_{u(T)}$. This behaviour which is dictated by Eq. (36) at once provides an explanation for the origin of a misconception: It is sometimes stated that non-Arrhenius kinetics in protein folding is limited to $k_{f(T)}$ while $k_{u(T)}$ usually follows Arrhenius-like kinetics.[36,49,53,61] It is readily apparent from these figures that if the experimental range of temperature over which the variation in $k_{u(T)}$ investigated is small, Arrhenius plots can appear to be linear (see Fig. 5A in Tan et al., 1996, Fig. 3 in Schindler and Schmid, 1996, and Fig. 6c in Jacob et al., 1999).[49,53,62] In fact, even if the temperature range is substantial, but owing to technical difficulties associated with measuring the unfolding rate constants below the freezing point of water, the range is restricted to temperatures above 273.16 K, $k_{u(T)}$ can still appear to be have a linear dependence on temperature in an Arrhenius plot since the curvature of the limbs in **Figures 10B, 11B** and **12B** is rather small. This can especially be the case if the number of experimental data points that define the Arrhenius plot is sparse. Consequently, the temperature-dependence of $k_{u(T)}$ can be fit equally well within statistical error to a linear function, and is apparent from inspection of the temperature-dependence of $k_{u(T)}$ of CI2 protein (see Fig. 4 in Tan et al., 1996).[62] Because $T_{H(TS-N)} \ll 273.16$ K for psychrophilic and mesophilic proteins, it is technically demanding, if not impossible, to experimentally demonstrate the increase in $k_{u(T)}$ for $T < T_{H(TS-N)}$ for the same. Nevertheless, the levelling-off of $k_{u(T)}$ at lower temperatures (see Fig. 3 in Schindler and Schmid, 1996), and extrapolation of data using non-linear fits (see Fig. 6B in Main et al., 1999) indicates this trend.[49,63] In principle, it may be possible to experimentally demonstrate this behaviour for those proteins whose $T_{H(TS-N)}$ is significantly above the freezing point of water. It is interesting to note that lattice models consisting of hydrophobic and polar residues (HP+ model) also capture this behaviour (see Fig. 22B in Chan and Dill, 1998).[50] As mentioned earlier, the cause of non-Arrhenius behaviour is a matter of some debate. However, because we have assumed a temperature-invariant prefactor and yet find that the kinetics are non-Arrhenius, it essentially implies that one does not need to invoke a *super-Arrhenius temperature-dependence of the configurational diffusion constant* to explain the non-Arrhenius behaviour of proteins.[17,36,48,50,55]

Once the temperature-dependence of $k_{f(T)}$ and $k_{u(T)}$ across a wide temperature range is known, the variation in the observed or the relaxation rate constant ($k_{obs(T)}$) with temperature may be readily ascertained using (see **Appendix**)[64]

$$\ln k_{obs(T)} = \ln \left[ k^0 \exp\left( -\frac{\alpha\left(\omega m_{D-N} - \sqrt{\varphi}\right)^2}{RT(\omega - \alpha)^2} \right) + k^0 \exp\left( -\frac{\omega\left(\sqrt{\varphi} - \alpha m_{D-N}\right)^2}{RT(\omega - \alpha)^2} \right) \right] \quad (37)$$

Inspection of **Figure supplement-1** of **Figures 10**, **11** and **12** demonstrates that $\ln(k_{obs(T)})$ vs temperature is a smooth "W-shaped" curve, with $k_{obs(T)}$ being dominated by $k_{f(T)}$ around $T_{H(TS-N)}$, and by $k_{u(T)}$ for $T < T_c$ and $T > T_m$, which is precisely why the kinks in $\ln(k_{obs(T)})$ occur around these temperatures. It is easy to see that at $T_c$ or $T_m$, $k_{f(T)} = k_{u(T)} \Rightarrow k_{obs(T)} = 2k_{f(T)} = 2k_{u(T)}$ and $\Delta G_{D-N(T)} = RT \ln\left(k_{f(T)}/k_{u(T)}\right) = 0$. In other words, for a two-state system, $T_c$ and $T_m$ measured at equilibrium must be identical to the temperatures at which $k_{f(T)}$ and $k_{u(T)}$ intersect.

This is a consequence of the *principle of microscopic reversibility*, i.e., the equilibrium and kinetic stabilities must be identical for a two-state system at all temperatures.[29]

Although a detailed discussion is beyond the scope of this article, the phenomenal increase in $k_{u(T)}$ and $k_{obs(T)}$ for $T < T_c$ and $T > T_m$ is due to the $\Delta G_{TS\text{-}N(T)}$ approaching zero (barrierless unfolding) at very low and high temperatures. Consequently, the unfolding rate constants approach their physical limit which is $k^0$; and any further decrease or an increase in temperature in the very low and high temperature regimes, respectively, must lead to a decrease in $k_{u(T)}$ (*Marcus-inversion*). This is readily apparent for FBP28 WW (**Figures 12B**, **Figure 12−figure supplement 1B** and **2B**). To summarise, for any fixed two-state folder, unfolding is *conventional barrier-limited* around $T = T_S$ and the position of the *curve-crossing* occurs in between the vertices of the DSE and the NSE parabolas. As the temperature deviates from $T_S$, $\Delta G_{TS\text{-}N(T)}$ decreases and eventually becomes zero at which point the *curve-crossing* occurs at the vertex of the NSE-parabola (i.e., the right arm of the DSE-parabola intersects the vertex of the NSE-parabola); and any further decrease or an increase in temperature in the very low and high temperature regimes, respectively, will cause unfolding to once again become barrier-limited with the *curve-crossing* occurring to the right of the vertex of the NSE-parabola (i.e., the right arm of the DSE-parabola intersects the right arm of the NSE-parabola).

Interestingly, in contrast to unfolding which can become barrierless at certain high and low temperature, folding is always barrier-limited with the absolute minimum of $\Delta G_{TS\text{-}D(T)}$ occurring when $T = T_S$; and any deviation in temperature from $T_S$ will only lead to an increase in $\Delta G_{TS\text{-}D(T)}$. Thus, from the perspective of parabolic hypothesis "*if folding is barrier-limited at $T_S$, then a two-state system at constant pressure and solvent conditions cannot spontaneously fold in a downhill manner, no matter what the temperature, and irrespective of whether or not it is an ultrafast folder.*" A corollary is that if there exists a chevron with a well-defined linear folding arm at $T_S$, then the prohibitive rule is that a two-state system at constant pressure and solvent conditions cannot spontaneously (i.e., unaided by co-solvents, ligands, metal ions etc.) fold by a downhill mechanism no matter what the temperature (see Popper, 1953, on why "prohibition" is as important as "confirming evidence" to any scientific method of inquiry). In other words, although the parabolic hypothesis predicts that *barrierless* and *Marcus-inverted regimes* for folding can occur, especially when $m_{D\text{-}N}$ is very small (**Figure 13**), the existence of a chevron with a well-defined linear folding arm at $T_S$ is sufficient to conclusively rule out such a scenario. It is imperative for the reader to take note of two aspects: First, the downhill folding scenario that is being referred to here is not the one wherein the denatured conformers fold to their native states *via* a *first-order* process with $k_{f(T)} \cong k^0$ (manifest when $\Delta G_{TS-D(T)}$ is approximately equal to ambient thermal noise, i.e., $\Delta G_{TS-D(T)} \cong 3RT$), but the controversial Type 0 scenario according to the Energy Landscape Theory, (see Fig. 6 in Onuchic et al., 1997) wherein the conformers in the DSE ostensibly reach the NSE without encountering any barrier ($\Delta G_{TS-D(T)} = 0$).[65-69] Second, the theoretical impossibility of a Type 0 scenario as claimed by the parabolic hypothesis comes with a condition and applies only for proteins that have linear folding chevron arms at $T_S$, and their folding proceeds without the need for accessory factors (metal ions for example) that are extrinsic with respect to the polypeptide chain. In other words, we are not outright ruling out a Type 0 scenario, since this could occur under certain conditions. However, what is being ruled out is that the proteins BBL and lambda repressor ($\lambda_{6-85}$) which have been touted to be paradigms for Type 0 scenario are most certainly not.[66,70,71] Further discussion on downhill scenarios is beyond the scope of this article and will be addressed elsewhere.

Comparison of the data shown in **Figures 10-12** and their figure supplements leads to an important conclusion: Just as sigmoidal changes in spectroscopic signals upon equilibrium chemical or thermal denaturation, Gaussian-like thermograms in differential scanning calorimetry (i.e., plots of change in partial heat capacity *vs* temperature), and the classic V-shaped chevrons (plots of $k_{obs(T)}$ *vs* chemical denaturant) are a characteristic feature of fixed two-state folders, so too must be the features of the temperature-dependences of $k_{obs(T)}$, $k_{f(T)}$ and $k_{u(T)}$. Although it might appear farfetched to arrive at this general conclusion merely from data on three proteins, the irrevocable requirement that $k_{f(T)}$ and $k_{u(T)}$ must approach each other as $T \rightarrow T_c$ or $T_m$, and that $k_{u(T)}$ must dominate $k_{f(T)}$ for $T < T_c$ and $T > T_m$ stems from the *principle of microscopic reversibility*, which unlike empirical laws, is grounded in statistical mechanics.[29] Consequently, the expectation is that $\ln(k_{u(T)})$ will have an approximate "V- or U-shape" and $\ln(k_{obs(T)})$ will have an approximate "W-shape" with respect to temperature (see Fig. 3 in Mayor et al., 2000, and Fig. 2 in Ghosh et al., 2007).[55,72] Further, since the large variation in equilibrium enthalpies and entropies of unfolding, including the pronounced curvature in $\Delta G_{D-N(T)}$ of proteins with temperature is due to the large and positive $\Delta C_{pD-N}$, a corollary is that "*non-Arrhenius kinetics can be particularly acute for reactions that are accompanied by large changes in the heat capacity.*" Because the change in heat capacity is proportional to the change in SASA, and since the change in SASA upon unfolding/folding increases with chain-length, "*non-Arrhenius kinetics, in general, can be particularly pronounced for large proteins, as compared to very small proteins and peptides.*" Now, Fersht and co-workers, by comparing the non-Arrhenius behaviour of the two-state-folding CI2 and the three-state-folding barnase argued that the pronounced curvature of $\ln(k_{f(T)}/T)$ of barnase as compared to CI2 in Eyring plots is a consequence of barnase folding *via* three-state kinetics (see Figs. 1 and 2 in Oliveberg et al., 1995).[48] Although there is no denying that barnase is not a fixed two-state system,[7,73] and their conclusion that "the non-Arrhenius behaviour of proteins is a consequence of the ensemble-averaged difference in heat capacity between the various reaction states" is rather remarkable, the pronounced curvature of barnase is highly unlikely to be a signature of three-state kinetics, but instead could be predominantly due to its larger size: While barnase is a 110-residue protein with $\Delta C_{pD-N}$ = 1.7 kcal.mol$^{-1}$.K$^{-1}$,[74] CI2 is significantly smaller in size (64 or 83 residues depending on the construct) with significantly lower $\Delta C_{pD-N}$ value (0.72 kcal.mol$^{-1}$.K$^{-1}$ for the short form and 0.79 kcal.mol$^{-1}$.K$^{-1}$ for the long form).[75] Although beyond the scope of this article and addressed elsewhere, it is important to recognize at this point that the non-linear temperature-dependence of equilibrium stability (see Fig.1 in Becktel and Schellman, 1987) is not the *cause* of non-Arrhenius kinetics, but instead is the *consequence* or the *equilibrium manifestation* of the underlying non-linear temperature-dependence of $k_{f(T)}$ and $k_{u(T)}$.

## IMPLICATIONS FOR PROTEIN FOLDING

The demonstration that the equilibrium stability and the rate constants are related to the mean and variance of the Gaussian distribution of the SASA of conformers in the DSE and NSE and the *curve-crossing* has certain implications for protein folding.[12]

First, analysing the effect of perturbations such as mutations on equilibrium stability purely in structural and native-centric terms, such as the removal or addition of certain interactions can be flawed because any perturbation that causes a change in the distribution of the conformers in either the DSE, or the NSE, or both, or the *curve-crossing* can cause a change in equilibrium stability. A corollary is that "*mutations need not be restricted to the structured elements of the native fold such as α-helices or β-sheets to cause a change in the rate* *constants or equilibrium stability, as compared to the wild type or the reference protein.*" Nagi and Regan's work offers a striking example: Increasing the loop-length using unstructured glycine linkers in the four-helix Rop1 leads to a dramatic change in the equilibrium stability, the rate constants, and the denaturant *m* value (i.e., the mean length of the RC), despite little or no effect on its native structure as determined by NMR and other spectroscopic probes, or its function as indicated by the ability of the mutants to form highly helical dimers and bind RNA (see Table 1 and Figs. 5 and 6 in Nagi and Regan, 1997).[76]

An important conclusion that we may draw from this Gaussian view of equilibrium stability is that if the DSEs and the NSEs of two related or unrelated spontaneously folding fixed two-state systems have identical mean SASAs and variances under identical environmental conditions (pressure, pH, temperature, ionic strength, co-solvents etc.), and if the position of the TSE along the SASA-RC is also identical for both, then irrespective of the: (*i*) primary sequence, including its length; (*ii*) amount of residual structure and the kinds of residual interactions in the DSE; (*iii*) topology of the native fold and the kinds of interactions that stabilize the native fold; and (*iv*) folding and unfolding rate constants, the said two-state systems must have identical equilibrium stabilities.

A further consequence of stability being a function of the variance of the distribution of the conformers in the ground states and the *curve-crossing* is that "*the contribution of a non-covalent interaction to equilibrium stability is not per se equal to the intrinsic Gibbs energy of the bond if the removal of the said interaction perturbs the variances of the both the DSE and NSE.*" What is implied by this is that, for example, if the removal of a salt-bridge in the native hydrophobic core of a hypothetical protein decreased the equilibrium stability by say 3 kcal.mol$^{-1}$, it is not logically incorrect to state that the removal of salt-bridge destabilized the protein by the said amount; however, what need not be true is the conclusion that the Gibbs energy of the said interaction is identical to the change in equilibrium stability brought forth by its removal.[77] It thus provides a rational explanation for why the Gibbs energies of molecular interactions such as hydrogen bonds and salt bridges as inferred from structural-perturbation-induced changes in equilibrium stability vary significantly across proteins.[78,79]

Second, since atoms are incompressible under conditions where the primary sequence exists as an entity, the maximum compaction (i.e., the reduction in SASA) that a given primary sequence can achieve is dictated by the excluded volume effect. Thus, the expectation is that as the chain length increases, the SASAs of the both the DSE and the NSE increase, albeit at different rates (otherwise $m_{D-N}$ and $\Delta C_{pD-N}$ will not increase with chain length); and on a decreasing absolute SASA scale, the position of the DSE and the NSE shift to the left with a concomitant increase in the relative separation between them. Conversely, a decrease in chain length will cause a shift in the position of both DSE and the NSE to the right with a concomitant decrease in the relative separation between them. What this implies is that the variance of the NSE is highly unlikely to be several orders of magnitude greater than that of the DSE (**Figures 6 and 7** and the values of the force constants given in the legends for Figures 10, 11 and 12). Thus, as long as the ratio of the variances of the DSE and the NSE is not a large number, and as long as there is a need to bury a minimum SASA by the denatured conformers for the microdomains (formed *en route* to the TSE or pre-existing in the DSE) to collide, coalesce and cross the Gibbs barrier to reach the native Gibbs basin, spontaneously-folding two-state systems can only be marginally stable.[80] A corollary of this is that "*the marginal stability of spontaneously-folding proteins is a consequence of physical chemistry.*"[81,82] In other words, intelligent life is built around marginal stability. This takes us back to Anfinsen's thermodynamic hypothesis:[83] Random mutations can lead to a repertoire of primary sequences *via* the central dogma; but whether or not these will fold into regular structures (spontaneously or not), and which when folded are stable enough to withstand thermal noise by virtue of their numerous intra-protein and protein-solvent interactions, and consequently, reside long enough in the native basin, thus giving rise to what we term "equilibrium stability" is ultimately governed by the laws of physical chemistry.[82,84,85] A detailed discussion on the physical basis for why the denatured conformers, in general, must diffuse a minimum distance along the SASA-RC for them to fold, and how this is related to the marginal stability of proteins is beyond the scope of this article and will be addressed in subsequent publications.

Third, it tells us that the equilibrium and kinetic behaviour of proteins *in vivo* can be significantly different from what we observe *in vitro*, not because the laws of physical chemistry do not apply to cellular conditions, but because the mean and variance of the distribution of the conformers in the ground states including the *curve-crossing*, owing to macromolecular crowding, can be significantly different *in vivo*.[86,87] This is apparent from the dramatic effect of metabolites such as glucose on the rate constants, the equilibrium stability, and $m_{D-N}$ (see Supplementary Fig. 4 in Wensley et al., 2010).[88] Thus, from the perspective of the parabolic hypothesis, isozymes are a consequence of a primary sequence optimization for function in a precisely defined environment. A further natural extension of folding and stability being functions of the Gaussian variances of the ground states and the *curve-crossing* is that the disulfide bonds in those proteins that fold in the highly crowded cytosol but must function in less crowded environments (for instance, cell-surface, soluble-secreted and extracellular matrix proteins) could be an evolutionary adaptation to fine tune the variances of the ground states for less crowded environments.[89] This will be dealt with in greater detail in subsequent publications where we will show from analysis of experimental data that the variances of DSE and the NSE are crucially dependent on their ensemble-averaged SASA, and the more expanded or solvent-exposed they are, the greater is their Gaussian variance, and *vice versa*.

Fourth, any experimental procedure that significantly perturbs the Gaussian mean and variance of the distribution of the SASA of the conformers in either the DSE or the NSE, or both, can significantly influence the outcome of the experiments, even if the final readout such as equilibrium stability is relatively unperturbed. These include treatments such as tethering the protein under investigation to a surface, or the covalent attachment of large donor and acceptor fluorophores such as those of the Alexa Fluor family (~1200 Dalton). Consequently, the conclusions based on data obtained from such measurements, although may be applicable to the system being studied, may not be readily extrapolated to the un-perturbed system that is either free in solution or devoid of extrinsic fluorophores.[90,91] This can especially be true if one places a large donor and acceptor labels on a very small protein or a peptide. A detailed comparison of the chemically-denatured and force-denatured DSEs of ubiquitin demonstrate that consistent with the large amount of data on the DSEs of proteins while the chemically-denatured DSE comprises significant population of α-helices, the force-denatured DSE is devoid of such secondary structural elements except under the lowest applied force.[92] Tethering a polypeptide to a surface has also been shown to greatly reduce the attempt frequency with which the protein samples its free energy.[93]

In summary, any perturbation− which may be intrinsic (*cis*-acting) or extrinsic (*trans*-acting) from the viewpoint of the primary sequence− that causes a change in the mean and the variance of the Gaussian distribution of the SASA of the conformers in the DSE, or the NSE, or both, or the *curve-crossing* can affect the equilibrium and kinetic behaviour of proteins. The *cis*-acting perturbations can be: (*i*) a change in the primary sequence *via* change in the gene sequence; (*ii*) any post-translation modification (phosphorylation, glycosylation, methylation, nitrosylation, acetylation, ubiquitylation, etc.) including covalent linking of fluorophores for the purpose of monitoring the dynamics of various protein conformational states; (*iii*) the introduction of disulfide bonds; and (*iv*) a change in the isotope composition of the primary sequence, i.e., homonuclear *vs* heteronuclear. The *trans*-acting perturbations can be: (*i*) a change in the temperature; (*ii*) a change in the pressure; (*iii*) a change in the solvent properties such as pH, ionic strength, solvent isotope, i.e., H$_2$O *vs* D$_2$O; (*iv*) macromolecular crowding; (*v*) selective non-covalent binding of any entity whether be it a small molecule (ligand-gating of ion channels, metal ions as in calcium signalling, metabolites, nucleotides and nucleotide-derivatives, the binding of substrates to enzymes, hormones, pharmaceutical drugs etc.), peptides and proteins (for example, a chaperone-client interaction, a chaperone-co-chaperone interaction, the cognate partner of an intrinsically denatured protein, the interaction between a G-protein coupled receptor and the G-protein, nucleotide exchange factors, protein and peptide therapeutics etc.), and DNA and RNA, to the conformers in either the DSE or the NSE; (*vi*) covalent tethering of the polypeptide to a surface, whether be it a synthetic such as a glass slide (typically employed in single-molecule experiments), or biologic such as the cell-walls of prokaryotes, the plasma membrane and the membranes of other organelles, such as those of the nucleus, the endoplasmic reticulum, the Golgi complex, mitochondria, chloroplasts, peroxisomes, lysosomes etc. (includes extrinsic, intrinsic, and membrane-spanning proteins); (*vii*) voltage, as in the case of voltage-gated ion channels; and (*viii*) molecular confinement, as in the case of chaperonin-assisted folding and proteasome-assisted degradation of proteins.

## METHODS

### Standard chevron-equation for two-state folding

The denaturant-dependence of the observed rate constant of any given two-state folder at constant temperature, pressure and solvent conditions (pH, buffer concentration, co-solvents other than the denaturant are constant) is given by the *standard chevron-equation*:[64]

$$\ln k_{\text{obs}(Den)(T)} = \ln\left[ k_{f(\text{H}_2\text{O})(T)} \exp\left(-m_{kf(T)}[Den]\right) + k_{u(\text{H}_2\text{O})(T)} \exp\left(m_{ku(T)}[Den]\right) \right] \quad (38)$$

$$\ln k_{\text{obs}(Den)(T)} = \ln\left[ k_{f(\text{H}_2\text{O})(T)} \exp\left(-\frac{m_{\text{TS-D}(T)}}{RT}[Den]\right) + k_{u(\text{H}_2\text{O})(T)} \exp\left(\frac{m_{\text{TS-N}(T)}}{RT}[Den]\right) \right] \quad (39)$$

where $k_{\text{obs}(Den)(T)}$ denotes the denaturant-dependence of the observed rate constant, $k_{f(\text{H}_2\text{O})(T)}$ and $k_{u(\text{H}_2\text{O})(T)}$ are the first-order rate constants for folding and unfolding, respectively, in water, [Den] is the denaturant concentration in molar, $m_{kf(T)}$ and $m_{ku(T)}$ are the denaturant dependencies of the natural logarithm of $k_{f(T)}$ and $k_{u(T)}$, respectively, with dimensions M$^{-1}$, $m_{TS-D(T)}$ ($=RT|m_{kf(T)}|$) and $m_{TS-N(T)}$ ($=RTm_{ku(T)}$) are parameters that are proportional to the ensemble-averaged difference in SASA between the DSE and TSE, and between the NSE and TSE, respectively, $R$ is the gas constant and $T$ is the absolute temperature (**Figure 1**).[25] Fitting $k_{obs(Den)(T)}$ *versus* [Den] data using non-linear regression to Eqs. (38) and (39) at constant temperature yields the said parameters. Conversely, if the values of $m_{TS-D(T)}$, $m_{TS-N(T)}$, $k_{f(H_2O)(T)}$ and $k_{u(H_2O)(T)}$ are known for any given two-state folder, one can readily simulate its chevron albeit without the experimental noise.

## Modified chevron-equation for two-state folding

The derivation of the *modified chevron-equation* is straightforward: $m_{TS-D(T)}$, $m_{TS-N(T)}$, $k_{f(H_2O)(T)}$ and $k_{u(H_2O)(T)}$ in Eq. (40) are replaced with Eqs. (9), (10), (11) and (12) respectively. The expanded equation is too long but the concise form is given by

$$\ln k_{obs(Den)(T)} = \ln\left[ k^0_{(H_2O)} \exp\left(-\frac{x}{RT}(\alpha x + [Den])\right) + k^0_{(H_2O)} \exp\left(\frac{y}{RT}([Den] - \omega y)\right) \right] \quad (40)$$

where $x = m_{TS-D(T)}$ and $y = m_{TS-N(T)}$. Fitting $k_{obs(Den)(T)}$ *versus* [Den] data to this equation using non-linear regression yields the Gibbs energy of unfolding in water, $\Delta G_{D-N(H_2O)(T)}$, $m_{D-N}$, the force constants $\alpha$ and $\omega$, and $k^0$. In the fitting procedure the statistical program starts the iterations with a pair of parabolas of arbitrary force constants and simultaneously: (*i*) adjusts the separation between their vertices along the abscissa such that it is exactly equal to $m_{D-N}$; (*ii*) adjusts the separation between their vertices along the ordinate such that it is exactly equal to $\Delta G_{D-N(T)}$; (*iii*) adjusts their curvature such that the separation between the *curve-crossing* and the vertex of the DSE-parabola along the abscissa is exactly equal to $m_{TS-D(T)}$, all the while looking for a suitable value of the prefactor such that: (*a*) $k_{f(T)}$ and $k_{u(T)}$ satisfy the Arrhenius equation at each one of the denaturant concentrations, and their sum is identical to the experimentally measured $k_{obs(T)}$ at that particular denaturant concentration; and (*b*) the principle of microscopic reversibility is satisfied at each one of the denaturant concentrations. The theory underlying the fitting procedure and its inherent limitations are addressed elsewhere.

## COMPETING FINANCIAL INTERESTS

The author declares no competing financial interests.

## COPYRIGHT INFORMATION

## APPENDIX

**Expression for the *curve-crossing* along the abscissa relative to the vertex of the DSE-Gibbs basin**

$$G_{\text{DSE}(r)(T)} = \alpha\left(0 - r_{(T)}\right)^2 = \alpha r_{(T)}^2 \tag{A1}$$

$$G_{\text{NSE}(r)(T)} = \omega\left(m_{\text{D-N}} - r_{(T)}\right)^2 - \Delta G_{\text{D-N}(T)} \tag{A2}$$

At the *curve-crossing* we have

$$G_{\text{DSE}(r\ddagger)(T)} = G_{\text{NSE}(r\ddagger)(T)} \Rightarrow \alpha r_{\ddagger(T)}^2 = \omega\left(m_{\text{D-N}} - r_{\ddagger(T)}\right)^2 - \Delta G_{\text{D-N}(T)} \tag{A3}$$

Expanding the term in the brackets and recasting gives

$$(\omega - \alpha) r_{\ddagger(T)}^2 - 2\omega m_{\text{D-N}} r_{\ddagger(T)} + \omega\left(m_{\text{D-N}}\right)^2 - \Delta G_{\text{D-N}(T)} = 0 \tag{A4}$$

The roots of this quadratic equation are given by

$$r_{\ddagger(T)} = \frac{-b \pm \sqrt{b^2 - 4ac}}{2a} \tag{A5}$$

Substituting the coefficients $a = (\omega - \alpha)$, $b = -2\omega m_{\text{D-N}}$ and $c = \omega\left(m_{\text{D-N}}\right)^2 - \Delta G_{\text{D-N}(T)}$ in Eq. (A5) and simplifying gives two options

$$r_{\ddagger(T)} = \frac{\omega m_{\text{D-N}} \pm \sqrt{\alpha\omega\left(m_{\text{D-N}}\right)^2 + \Delta G_{\text{D-N}(T)}(\omega - \alpha)}}{(\omega - \alpha)} \tag{A6}$$

The point where the right arm of the DSE-parabola intersects the left arm of the NSE-parabola along the RC is given by

$$r_{\ddagger(T)} = \frac{\omega m_{D-N} - \sqrt{\alpha\omega(m_{D-N})^2 + \Delta G_{D-N(T)}(\omega-\alpha)}}{(\omega-\alpha)} \quad (A7)$$

The point where the right arm of the DSE-parabola intersects the right arm of the NSE-parabola along the RC is given by

$$r_{\ddagger(T)} = \frac{\omega m_{D-N} + \sqrt{\alpha\omega(m_{D-N})^2 + \Delta G_{D-N(T)}(\omega-\alpha)}}{(\omega-\alpha)} \quad (A8)$$

Because the TSE occurs in between the vertices of the DSE and the NSE Gibbs basins along the abscissa (this is not always true and is addressed in subsequent publications), we ignore Eq. (A8). Substituting $\lambda = \alpha(m_{D-N})^2$ in Eq. (A7) gives

$$r_{\ddagger(T)} \equiv m_{TS-D(T)} = \frac{\omega m_{D-N} - \sqrt{\lambda\omega + \Delta G_{D-N(T)}(\omega-\alpha)}}{(\omega-\alpha)} \quad (A9)$$

Substituting $\varphi = \lambda\omega + \Delta G_{D-N(T)}(\omega-\alpha)$ in Eq. (A9) yields the final form

$$m_{TS-D(T)} = \frac{\omega m_{D-N} - \sqrt{\varphi}}{(\omega-\alpha)} \quad (A10)$$

**Expression for the *curve-crossing* along the abscissa relative to the vertex of the NSE-Gibbs basin**

For a two-state folder we have

$$m_{TS-N(T)} = m_{D-N} - m_{TS-D(T)} \quad (A11)$$

Substituting Eq. (A10) in (A11) gives

$$m_{TS-N(T)} = m_{D-N} - \frac{\omega m_{D-N} - \sqrt{\varphi}}{(\omega-\alpha)} = \frac{(m_{D-N}(\omega-\alpha)) - (\omega m_{D-N} - \sqrt{\varphi})}{(\omega-\alpha)} \quad (A12)$$

Simplifying Eq. (A12) gives

$$m_{\text{TS-N}(T)} = \frac{\omega m_{\text{D-N}} - \alpha m_{\text{D-N}} - \omega m_{\text{D-N}} + \sqrt{\varphi}}{(\omega - \alpha)} = \frac{\sqrt{\varphi} - \alpha m_{\text{D-N}}}{(\omega - \alpha)} \tag{A13}$$

## Expression for $k_{u(T)}$ using the principle of microscopic reversibility

For a two-state folder, we have from the principle of microscopic reversibility[29]

$$\Delta G_{\text{D-N}(T)} = -RT \ln\left(\frac{k_{u(T)}}{k_{f(T)}}\right) \Rightarrow k_{u(T)} = k_{f(T)} \exp\left(-\frac{\Delta G_{\text{D-N}(T)}}{RT}\right) \tag{A14}$$

Substituting Eq. (11) in (A14) and simplifying gives

$$k_{u(T)} = k^0 \exp\left(-\frac{(\omega - \alpha)^2 \Delta G_{\text{D-N}(T)} + \alpha\left(\omega m_{\text{D-N}} - \sqrt{\varphi}\right)^2}{RT(\omega - \alpha)^2}\right) \tag{A15}$$

## Expressions for the Gibbs barrier heights and the rate constants in terms of $\beta_{\text{T(fold)}(T)}$ and $\beta_{\text{T(unfold)}(T)}$

We have from Tanford's adaptation of the Brønsted framework to solvent denaturation of proteins[25]

$$\frac{m_{\text{TS-D}(T)}}{m_{\text{D-N}}} = \beta_{\text{T(fold)}(T)} \Rightarrow \left(m_{\text{TS-D}(T)}\right)^2 = \left(\beta_{\text{T(fold)}(T)} m_{\text{D-N}}\right)^2 \tag{A16}$$

$$\frac{m_{\text{TS-N}(T)}}{m_{\text{D-N}}} = \beta_{\text{T(unfold)}(T)} \Rightarrow \left(m_{\text{TS-N}(T)}\right)^2 = \left(\beta_{\text{T(unfold)}(T)} m_{\text{D-N}}\right)^2 \tag{A17}$$

Substituting Eq. (A16) in (1) and Eq. (A17) in (2) yield

$$\Delta G_{\text{TS-D}(T)} = \alpha\left(\beta_{\text{T(fold)}(T)} m_{\text{D-N}}\right)^2 = \lambda \beta_{\text{T(fold)}(T)}^2 \tag{A18}$$

$$\Delta G_{\text{TS-N}(T)} = \omega\left(m_{\text{TS-N}(T)}\right)^2 = \omega\left(\beta_{\text{T(unfold)}(T)} m_{\text{D-N}}\right)^2 = \frac{\lambda \omega \beta_{\text{T(unfold)}(T)}^2}{\alpha} \tag{A19}$$

Substituting Eq. (A18) in (11) and (A19) in (12) yield

$$k_{f(T)} = k^0 \exp\left(-\frac{\alpha\left(\beta_{\text{T(fold)}(T)} m_{\text{D-N}}\right)^2}{RT}\right) = k^0 \exp\left(-\frac{\lambda\beta_{\text{T(fold)}(T)}^2}{RT}\right) \quad (A20)$$

$$k_{u(T)} = k^0 \exp\left(-\frac{\omega\left(\beta_{\text{T(unfold)}(T)} m_{\text{D-N}}\right)^2}{RT}\right) = k^0 \exp\left(-\frac{\lambda\omega\beta_{\text{T(unfold)}(T)}^2}{\alpha RT}\right) \quad (A21)$$

**Expressions for the *curve-crossing*, $k_{f(T)}$ and $k_{u(T)}$ at the midpoint of cold and heat denaturation**

At the midpoint of thermal ($T_m$) or cold denaturation ($T_c$), $\Delta G_{\text{D-N}(T_m/T_c)} = 0$. Consequently, Eqs. (A10) and (A13) become

$$m_{\text{TS-D}(T)}\Big|_{T=T_m,T_c} = \frac{\omega m_{\text{D-N}} - \sqrt{\alpha\omega\left(m_{\text{D-N}}\right)^2}}{(\omega - \alpha)} = \frac{m_{\text{D-N}}\left(\omega - \sqrt{\alpha\omega}\right)}{(\omega - \alpha)} \quad (A22)$$

$$m_{\text{TS-N}(T)}\Big|_{T=T_m,T_c} = \frac{\sqrt{\alpha\omega\left(m_{\text{D-N}}\right)^2} - \alpha m_{\text{D-N}}}{(\omega - \alpha)} = \frac{m_{\text{D-N}}\left(\sqrt{\alpha\omega} - \alpha\right)}{(\omega - \alpha)} \quad (A23)$$

Eqs. (A22) and (A23) may be recast in terms of $\beta_{\text{T(fold)}(T)}$ and $\beta_{\text{T(unfold)}(T)}$ to give

$$\beta_{\text{T(fold)}(T)}\Big|_{T=T_m,T_c} = \frac{m_{\text{TS-D}(T)}}{m_{\text{D-N}}}\Big|_{T=T_m,T_c} = \frac{\omega - \sqrt{\alpha\omega}}{(\omega - \alpha)} \quad (A24)$$

$$\beta_{\text{T(unfold)}(T)}\Big|_{T=T_m,T_c} = \frac{m_{\text{TS-N}(T)}}{m_{\text{D-N}}}\Big|_{T=T_m,T_c} = \frac{\sqrt{\alpha\omega} - \alpha}{(\omega - \alpha)} \quad (A25)$$

Substituting Eqs. (A24) and (A25) in (13) and (14), respectively, and simplifying gives expressions for the rate constants for folding and unfolding at $T_m$ or $T_c$

$$k_{f(T)}\Big|_{T=T_m,T_c} = k_{u(T)}\Big|_{T=T_m,T_c} = k^0 \exp\left(-\frac{\lambda\omega\left(\omega + \alpha - 2\sqrt{\alpha\omega}\right)}{RT(\omega - \alpha)^2}\right)\Bigg|_{T=T_m,T_c} \quad (A26)$$

Eqs. (A22) − (A26) demonstrate that the *curve-crossing*, $k_{f(T)}$ and $k_{u(T)}$, $\beta_{\text{T(fold)}(T)}$ and $\beta_{\text{T(unfold)}(T)}$ at $T_m$ or $T_c$ for any given two-state system are defining constants when solvent and pressure are defined since they depend only on the length of the RC and the force constants, all of which are invariant with temperature. Consequently, these are properties that are dependent purely on the primary sequence when pressure and solvent are defined. There are other defining relationships at $T_m$ or $T_c$. Substituting $\Delta G_{D\text{-}N(T_m/T_c)} = 0$ in Eq. (3) gives

$$\omega \left( m_{\text{TS-N}(T)} \right)^2 \bigg|_{T=T_m,T_c} = \alpha \left( m_{\text{TS-D}(T)} \right)^2 \bigg|_{T=T_m,T_c} \Rightarrow \frac{m_{\text{TS-D}(T)}}{m_{\text{TS-N}(T)}} \bigg|_{T=T_m,T_c} = \sqrt{\frac{\omega}{\alpha}} \bigg|_{T=T_m,T_c} \qquad (A27)$$

Recasting Eq. (A27) in terms of Eqs. (23) and (24) gives

$$\frac{m_{\text{TS-D}(T)}}{m_{\text{TS-N}(T)}} \bigg|_{T=T_m,T_c} = \sqrt{\frac{\sigma^2_{\text{DSE}(T)}}{\sigma^2_{\text{NSE}(T)}}} \bigg|_{T=T_m,T_c} = \frac{\sigma_{\text{DSE}(T)}}{\sigma_{\text{NSE}(T)}} \bigg|_{T=T_m,T_c} \qquad (A28)$$

Eqs. (A27) and (A28) demonstrate that at $T_m$ or $T_c$, the ratio of the slopes of the folding and unfolding arms of the chevron, or the ratio of the distances by which the conformers in the DSE and NSE travel from the mean of their respective ensembles to reach the TSE along the $m_{D\text{-}N}$ RC for a given two-state system is identical to: (*i*) the square root of the ratio of the force constants of the NSE and the DSE; or (*ii*) the ratio of the standard deviations of the DSE and NSE Gaussians. A corollary is that irrespective of the primary sequence, or the topology of the native state, or the residual structure in the DSE, if for a spontaneously folding two-state system at constant pressure and solvent conditions it is found that at a certain temperature the ratio of the distances by which the denatured and the native conformers must travel from the mean of their ensemble to reach the TSE along the SASA RC is identical to the ratio of the standard deviations of the Gaussian distribution of the SASA of the conformers in the DSE and the NSE, then at this temperature the Gibbs energy of unfolding or folding must be zero.

**Expression for the temperature-dependence of the observed rate constant**

The observed rate constant $k_{\text{obs}(T)}$ for a two-state system is the sum of $k_{f(T)}$ and $k_{u(T)}$.[64] Therefore, we can write

$$k_{\text{obs}(T)} = k_{f(T)} + k_{u(T)} \Rightarrow \ln k_{\text{obs}(T)} = \ln \left( k_{f(T)} + k_{u(T)} \right) \qquad (A29)$$

Substituting Eqs. (11) and (12) in (A29) gives

$$\ln k_{\text{obs}(T)} = \ln\left[k^0 \exp\left(-\frac{\alpha\left(\omega m_{\text{D-N}} - \sqrt{\varphi}\right)^2}{RT(\omega-\alpha)^2}\right) + k^0 \exp\left(-\frac{\omega\left(\sqrt{\varphi} - \alpha m_{\text{D-N}}\right)^2}{RT(\omega-\alpha)^2}\right)\right] \quad \text{(A30)}$$

# FIGURES

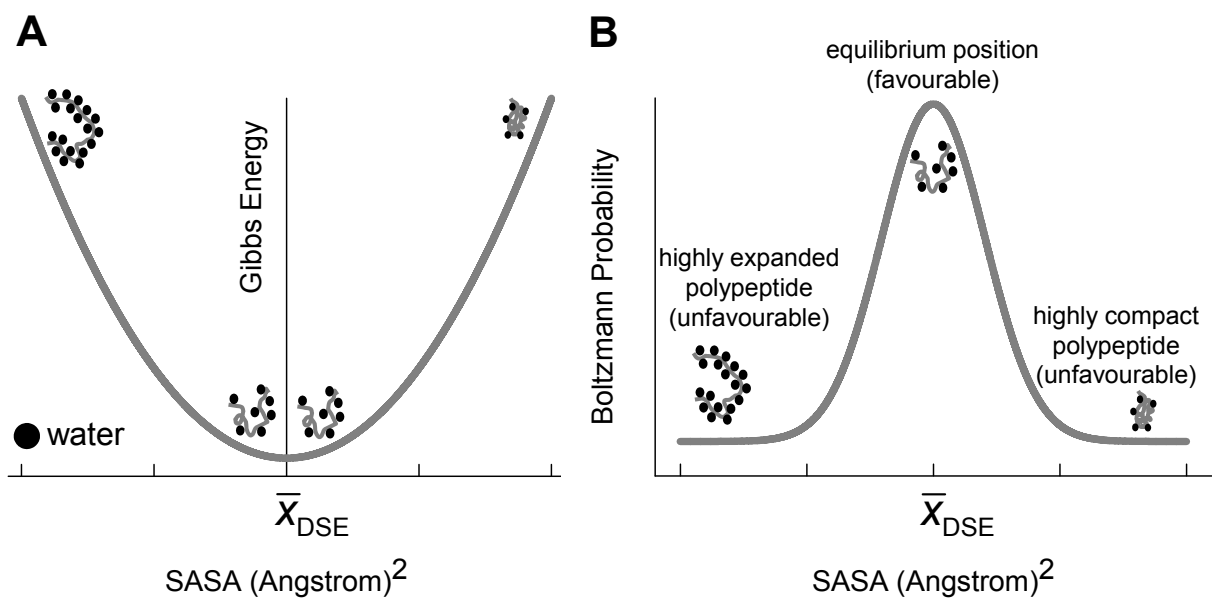

**Figure 1.**

**Gibbs energies and Boltzmann probabilities for the conformers in the DSE of a two-state system at equilibrium under folding conditions.**

**(A)** The Gibbs energies of the conformers in the DSE are proportional to the square of the displacement from the mean SASA of the ensemble. **(B)** Boltzmann probabilities for the conformers in the DSE. For any given temperature, pressure and solvent conditions, the most probable microstates are those whose Gibbs energies are the least.

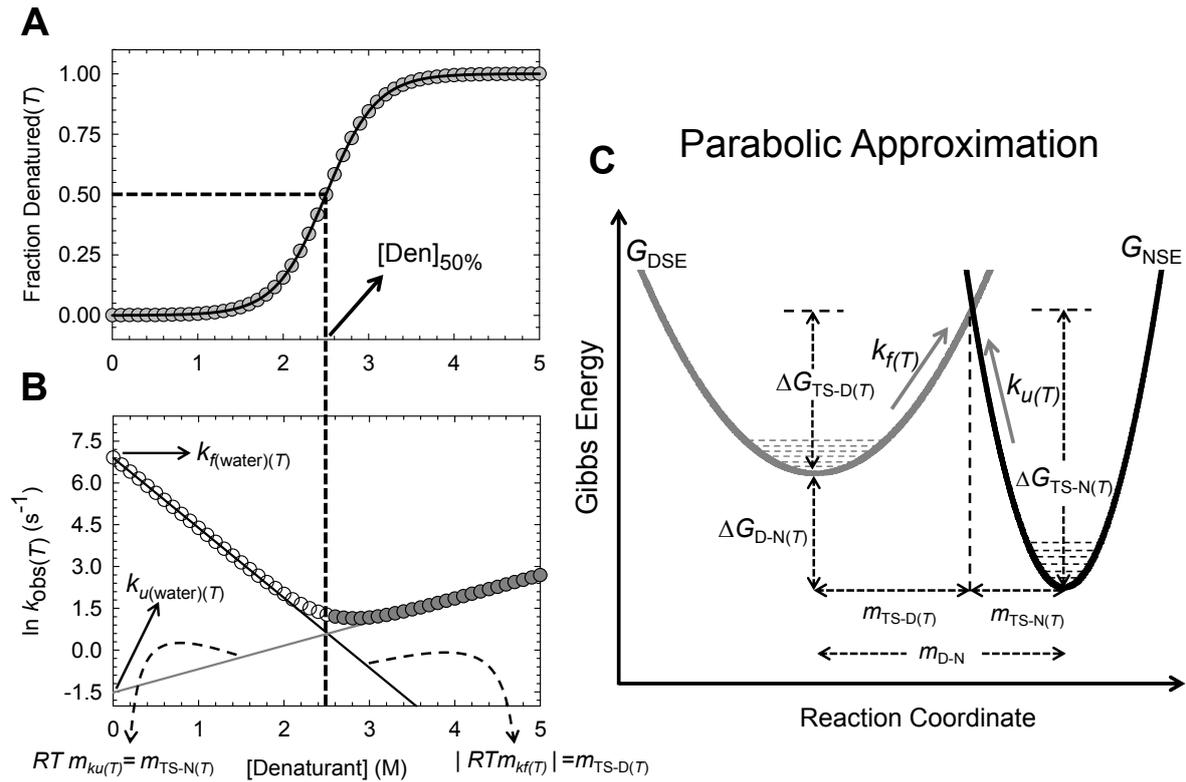

**Figure 2.**

**Standard equilibrium and kinetic parameters from chemical denaturation and their relationship to parabolic Gibbs energy curves.**

**(A)** An equilibrium chemical denaturation curve simulated using standard two-state equations for a hypothetical two-state protein with $\Delta G_{D-N(T)}$ = 5 kcal.mol$^{-1}$; $m_{D-N}$ = 2 kcal.mol$^{-1}$.M$^{-1}$; [Den$_{50\%}$] = 2.5 M and $T$ = 298.16 K.[64] The midpoint of chemical denaturation is given by [Den$_{50\%}$]. **(B)** A corresponding chevron simulated using Eq. (38) with $k_{f(\text{water})(T)}$ = 1000 s$^{-1}$; $k_{u(\text{water})(T)}$ = 0.216 s$^{-1}$; $m_{TS-D(T)}$ and $m_{TS-N(T)}$ are 1.5 and 0.5 kcal.mol$^{-1}$.M$^{-1}$, respectively, and $T$ = 298.16 K. The denaturant-dependences of ln $k_{f(T)}$ and ln $k_{u(T)}$ are given by $m_{kf(T)}$ (solid black line) and $m_{ku(T)}$ (solid grey line), respectively. **(C)** Parabolic approximation for a hypothetical two-state protein. The parabolas were generated according to Eqs. (4) and (5). The Gibbs barrier heights for folding and unfolding are given by $\Delta G_{TS-D(T)}$ and $\Delta G_{TS-N(T)}$, respectively. The temperature-invariant mean length of the RC is given by $m_{D-N}$, the mean position of the TSE with respect to the DSE and the NSE along the RC is given by $m_{TS-D(T)}$ and $m_{TS-N(T)}$, respectively.

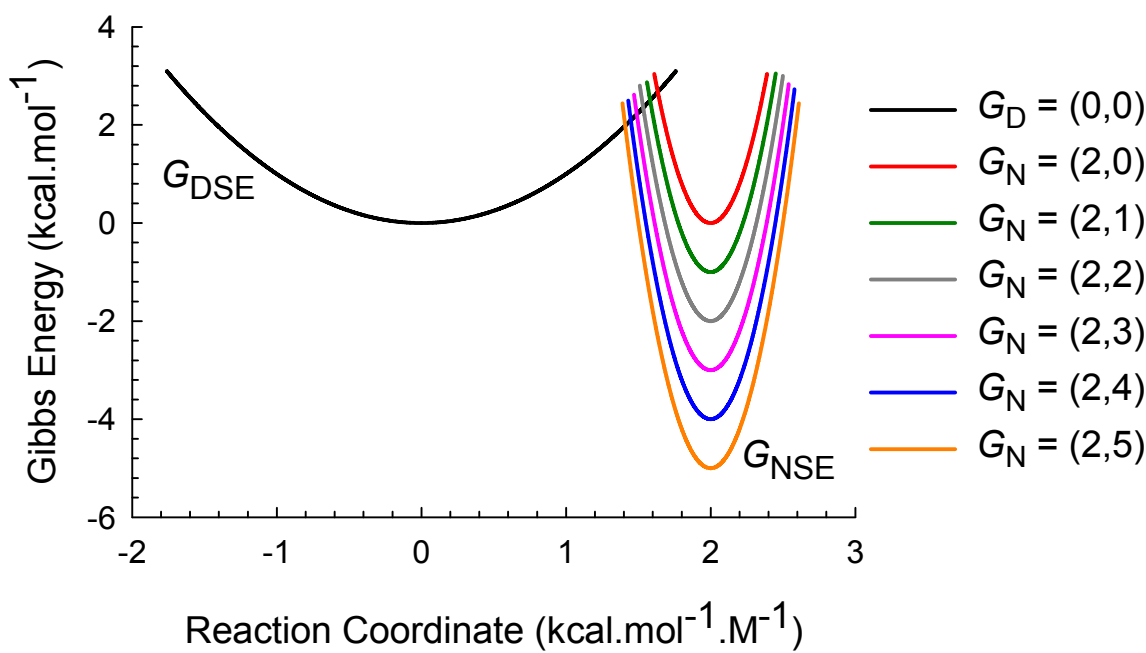

**Figure 3.**

**The effect of temperature-induced changes in $\Delta G_{\text{D-N}(T)}$ on the energetics and placement of the TSE along the RC.**

Gibbs energy curves according to Eqs. (4) and (5) for a hypothetical two-state system with $\alpha = 1$ $M^2 \cdot \text{mol} \cdot \text{kcal}^{-1}$, $\omega = 20$ $M^2 \cdot \text{mol} \cdot \text{kcal}^{-1}$, and $m_{\text{D-N}} = 2$ $\text{kcal} \cdot \text{mol}^{-1} \cdot M^{-1}$. The coordinates of the vertices of the parabolas are given in the legend. $\Delta G_{\text{D-N}(T)}$ is a maximum at $T_S$ (5 $\text{kcal} \cdot \text{mol}^{-1}$, orange curve) and is 0 $\text{kcal} \cdot \text{mol}^{-1}$ at $T_m$ (red curve). As the protein is increasingly destabilized the position of the *curve-crossing* along the abscissa shifts towards the vertex of the NSE-parabola (Hammond movement).[94] Note that $\Delta G_{\text{TS-D}(T)}$ and $\Delta G_{\text{TS-N}(T)}$ are a minimum and a maximum, respectively, at $T_S$. Increasing the temperature from $T_S \rightarrow T_m$ leads to an increase in $\Delta G_{\text{TS-D}(T)}$ and a concomitant decrease in $\Delta G_{\text{TS-N}(T)}$. Although the change in stability is shown relative to the DSE, a decrease in stability can be due to a stabilized DSE or destabilized NSE or both. Conversely, an increase in stability can be due to a destabilized DSE or a stabilized NSE or both.

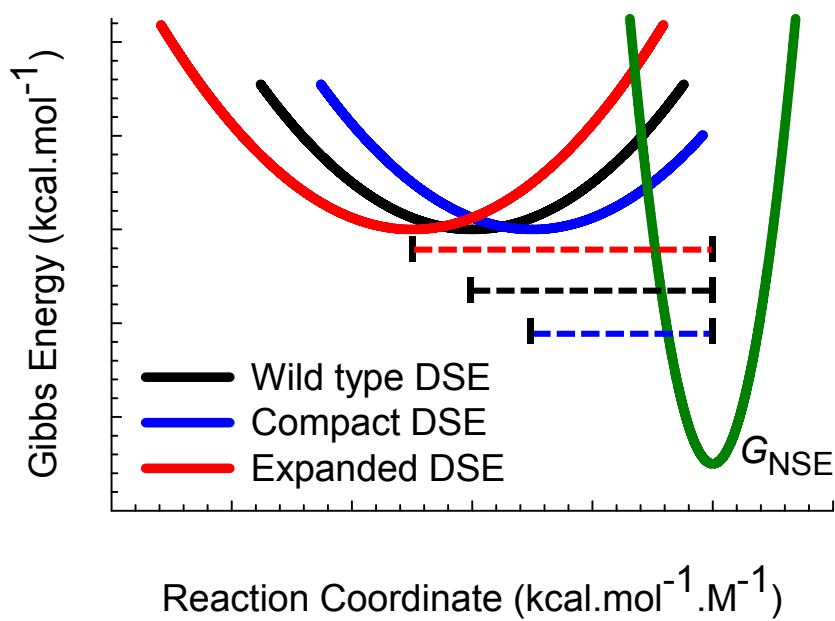

**Figure 4.**

**The effect of perturbation-induced change in the mean length of the RC on the height of the Gibbs barrier to folding/unfolding.**

Gibbs energy curves according to Eqs. (4) and (5) for a hypothetical two-state system with α = 1 $M^2.mol.kcal^{-1}$ and ω = 20 $M^2.mol.kcal^{-1}$. The dotted lines denote $m_{D-N}$ for the wild type and its mutants. A perturbation-induced (mutation/co-solvents/pH) increase in the mean length of the RC as compared to the wild type/reference protein state can lead to an increase in $\Delta G_{TS-D(T)}$ and $\Delta G_{TS-N(T)}$. This can manifest as a significant and simultaneous decrease in $k_{f(T)}$ and $k_{u(T)}$. Conversely, a contraction of the RC can cause a decrease $\Delta G_{TS-D(T)}$ and $\Delta G_{TS-N(T)}$, leading to a simultaneous increase in $k_{f(T)}$ and $k_{u(T)}$. These effects can sometimes manifest as an inverse linear correlation between ln $k_{f(T)}$ and $m_{TS-D(T)}$.[19] Comparison of Figures 3 and 4 shows that a unit change in $m_{D-N}$, in general, causes a much larger change in $\Delta G_{TS-D(T)}$ and $\Delta G_{TS-N(T)}$ than a unit change in $\Delta G_{D-N(T)}$. Nevertheless, this is an oversimplification since perturbations that cause a change in $m_{D-N}$ usually lead to concomitant changes in the curvature of the parabolas and the prefactor (addressed in subsequent publications).

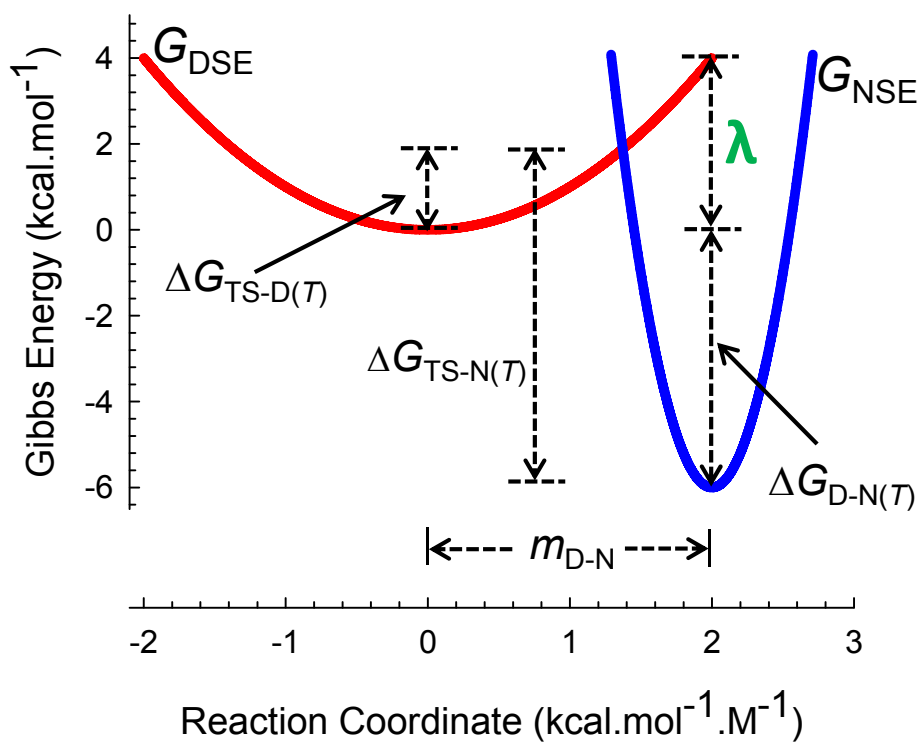

**Figure 5.**

**Marcus reorganization energy (λ) for two-state protein folding.**

Parabolic Gibbs energy curves according to Eqs. (4) and (5) for a hypothetical two-state folder with force constants $\alpha$ = 1 M$^2$.mol.kcal$^{-1}$, $\omega$ = 20 M$^2$.mol.kcal$^{-1}$, $\Delta G_{D-N(T)}$ = 6 kcal.mol$^{-1}$, $m_{D-N}$ = 2 kcal.mol$^{-1}$.M$^{-1}$ and $\lambda$ = 4 kcal.mol$^{-1}$.

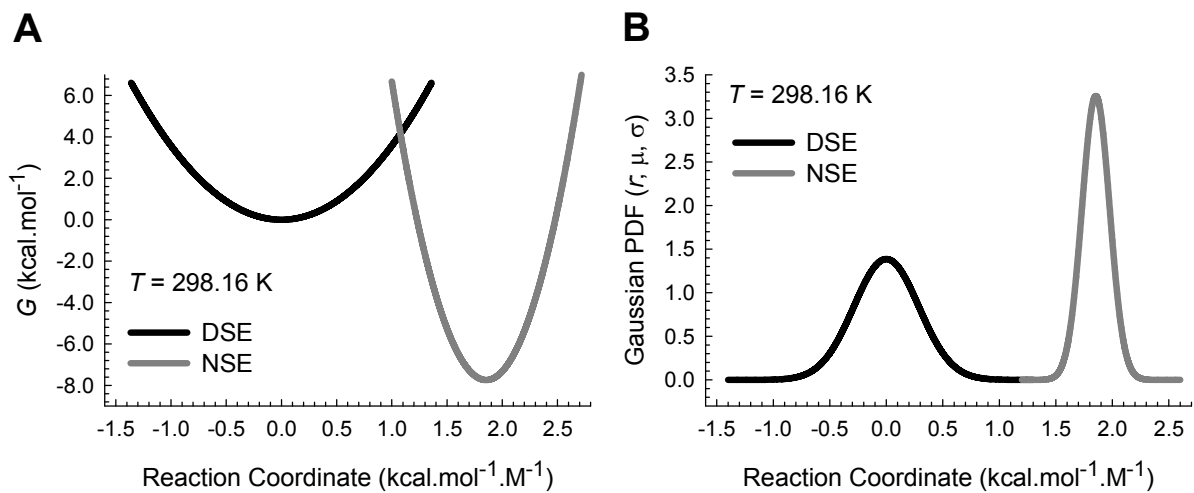

**Figure 6.**

**Correspondence between Gibbs parabolas and the Gaussian PDFs for the wild type 64-residue CI2.**

**(A)** Parabolic Gibbs energy curves according to Eqs. (4) and (5) with $\alpha = 3.576$ $M^2.mol.kcal^{-1}$ and $\omega = 19.759$ $M^2.mol.kcal^{-1}$, $m_{D-N} = 1.8543$ $kcal.mol^{-1}.M^{-1}$ and $\Delta G_{D-N(T)} = 7.7442$ $kcal.mol^{-1}$. The separation between the vertex of the DSE-parabola and the *curve-crossing* ($m_{TS-D(T)}$) is 1.0782 $kcal.mol^{-1}.M^{-1}$, and between the vertex of the NSE-parabola and the *curve-crossing* ($m_{TS-N(T)}$) is 0.7761 $kcal.mol^{-1}.M^{-1}$. The absolute values of $\Delta G_{TS-D(T)}$ and $\Delta G_{TS-N(T)}$ are 4.1571 $kcal.mol^{-1}$ and 11.9014 $kcal.mol^{-1}$, respectively. The values of the force constants were obtained by fitting the chevron of the wild type CI2 to a modified chevron-equation (see **Methods**). The parameters required to generate the chevron were taken from Tables 2 and 3 of Itzhaki et al., 1995.[95] **(B)** Gaussian PDFs for the DSE and NSE generated using $p(r) = \frac{1}{\sqrt{2\pi\sigma^2}} \exp\left(-(r-\mu)^2/2\sigma^2\right)$, where $r$ is any point on the abscissa, $\mu = 0$ $kcal.mol^{-1}.M^{-1}$ and $\sigma^2 = 0.0828$ $kcal^2.mol^{-2}.M^{-2}$ for the DSE-Gaussian, and $\mu = 1.8543$ $kcal.mol^{-1}.M^{-1}$ and $\sigma^2 = 0.0149$ $kcal^2.mol^{-2}.M^{-2}$ for the NSE-Gaussian. The area enclosed by the DSE and NSE-Gaussians is unity. The experimental conditions are as follows: 50 mM Mes, pH 6.25, 298.16 K.[95]

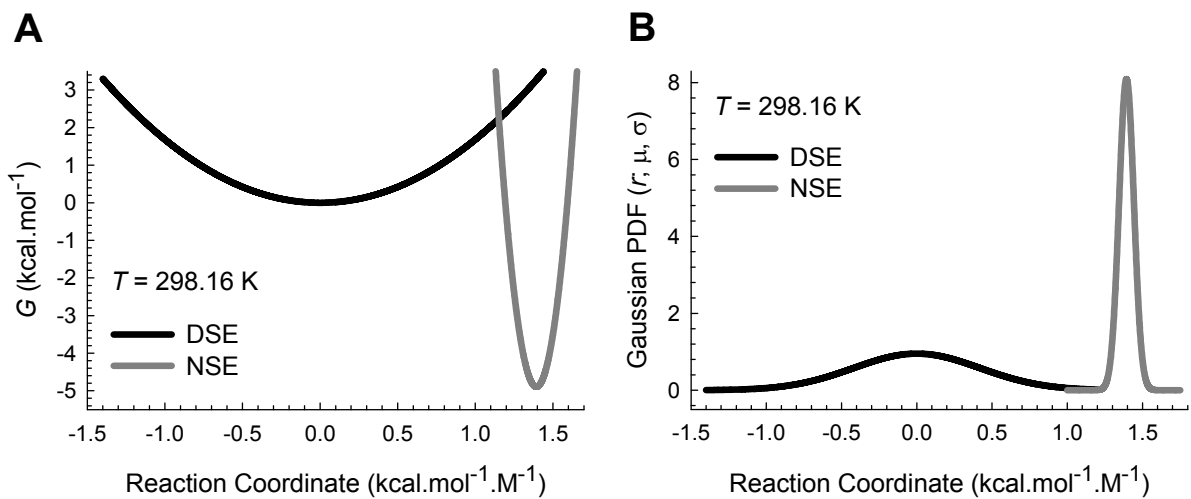

**Figure 7**

**Correspondence between Gibbs parabolas and Gaussian PDFs for the B domain of Staphylococcal protein A (BdpA Y15W).**

**(A)** Parabolic Gibbs energy curves according to Eqs. (4) and (5) with $\alpha = 1.683$ $M^2.mol.kcal^{-1}$ and $\omega = 122.175$ $M^2.mol.kcal^{-1}$, $m_{D-N} = 1.3936$ $kcal.mol^{-1}.M^{-1}$ and $\Delta G_{D-N(T)} = 4.8979$ $kcal.mol^{-1}$. The separation between the vertex of the DSE-parabola and the *curve-crossing* is 1.152 $kcal.mol^{-1}.M^{-1}$ ($m_{TS-D(T)}$), and between the *curve-crossing* and the vertex of the NSE-parabola is 0.2416 $kcal.mol^{-1}.M^{-1}$ ($m_{TS-N(T)}$). The absolute values of $\Delta G_{TS-D(T)}$ and $\Delta G_{TS-N(T)}$ are 2.2335 $kcal.mol^{-1}$ and 7.1314 $kcal.mol^{-1}$, respectively. The values of the force constants were obtained by fitting the chevron of BdpA Y15W to a modified chevron-equation (see **Methods**). The parameters required to generate the chevron were taken from Table 7 of Sato and Fersht, 2007.[58] **(B)** Gaussian PDFs for the DSE and NSE generated using $p(r) = \frac{1}{\sqrt{2\pi\sigma^2}}\exp(-(r-\mu)^2/2\sigma^2)$, where $r$ is any point on the abscissa, $\mu = 0$ $kcal.mol^{-1}.M^{-1}$ and $\sigma^2 = 0.1760$ $kcal^2.mol^{-2}.M^{-2}$ for the DSE-Gaussian, and $\mu = 1.3936$ $kcal.mol^{-1}.M^{-1}$ and $\sigma^2 = 0.002424$ $kcal^2.mol^{-2}.M^{-2}$ for the NSE-Gaussian. The area under the Gaussians is unity. The experimental conditions are as follows: 50 mM sodium acetate, 100 mM NaCl, pH 5.5, 298.16 K.[58]

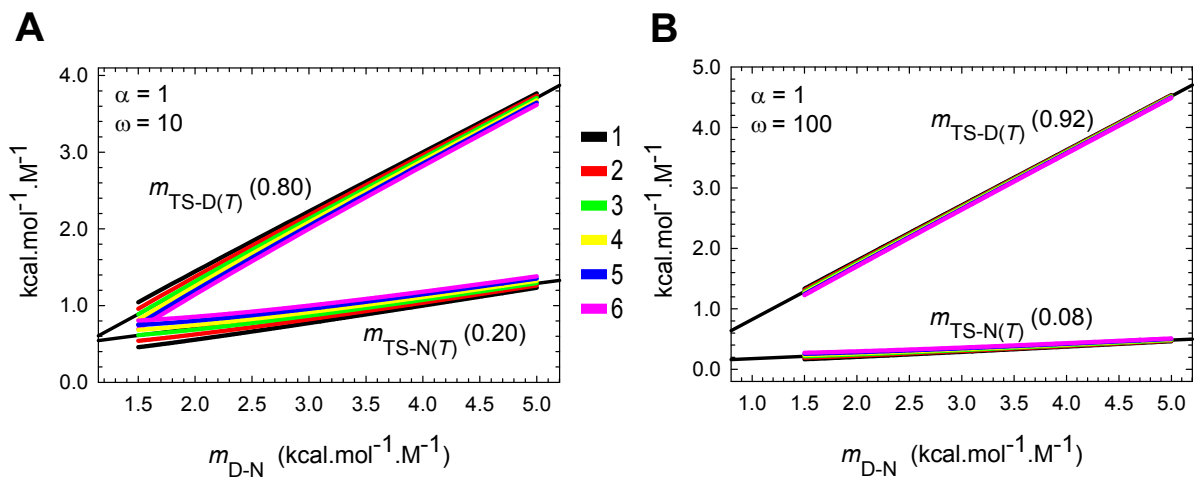

**Figure 8.**

**Predicted dependence of $m_{TS-D(T)}$ and $m_{TS-N(T)}$ on $m_{D-N}$ for two hypothetical two-state folders according to Eqs. (9) and (10).**

**(A)** Variation in $m_{TS-D(T)}$ and $m_{TS-N(T)}$ with $m_{D-N}$ for a protein with $\alpha = 1$ and $\omega = 10$ M$^2$.mol.kcal$^{-1}$. **(B)** Variation in $m_{TS-D(T)}$ and $m_{TS-N(T)}$ with $m_{D-N}$ for a protein with $\alpha = 1$ and $\omega = 100$ M$^2$.mol.kcal$^{-1}$. For a given equilibrium stability (i.e., separation between the vertices of the DSE and NSE-parabolas along the ordinate), if the separation between the vertices of the DSE and NSE-parabolas along the abscissa is varied (as in **Figure 4**), a large fraction of the variation in $m_{D-N}$ is manifest in $m_{TS-D(T)}$ and not $m_{TS-N(T)}$; and this is particularly pronounced for systems with high $\beta_{T(fold)(T)}$, i.e., $\omega \gg \alpha$. The linear fits are aggregate slopes of the variation in $m_{TS-D(T)}$ and $m_{TS-N(T)}$ (shown as insets) for the six sub-systems with equilibrium stabilities ranging from 1 to 6 kcal.mol$^{-1}$ and indicated by the legend (common to both the plots). This behaviour is precisely what is observed for real proteins (**Figure 9**). Note that because $\beta_{T(fold)(T)} + \beta_{T(unfold)(T)} = 1$ for two-state systems, the sum of slopes of $m_{TS-D(T)}$ and $m_{TS-N(T)}$ must be unity.

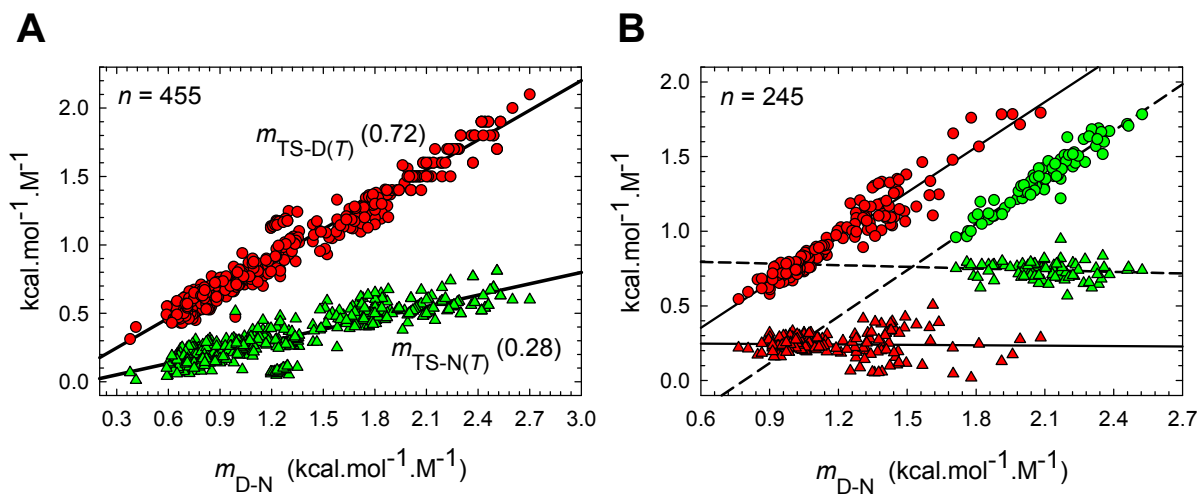

**Figure 9.**

**Experimentally determined mutation-induced variation in $m_{TS-D(T)}$ and $m_{TS-N(T)}$ with $m_{D-N}$ for a total of 700 mutants from 18 two-state systems.**

**(A)** Approximately 72% of a unit change in $m_{D-N}$ is manifest in $m_{TS-D(T)}$ (R = 0.98) and the rest in $m_{TS-N(T)}$ (R = 0.87), for a total of 455 mutants from 12 two-state systems. **(B)** Almost all of the change in $m_{D-N}$ is restricted to $m_{TS-D(T)}$ leaving $m_{TS-N(T)}$ virtually unchanged for a total of 245 mutants from 6 two-state systems. The data appear as two subsets (red and green symbols) owing to the large differences in their $m_{D-N}$ values. The slope and correlation coefficient for the: (*i*) red circles are 1.00 and 0.96, respectively; (*ii*) red triangles are -0.0089 and 0.03, respectively; (*iii*) green circles are 1.04 and 0.96, respectively; and (*iv*) green triangles are -0.04 and 0.12, respectively. As stipulated by theory, the sum of the slopes of the linear regression is unity. See also Figures 7 and 9 in Sanchez and Kiefhaber, 2003.[19]

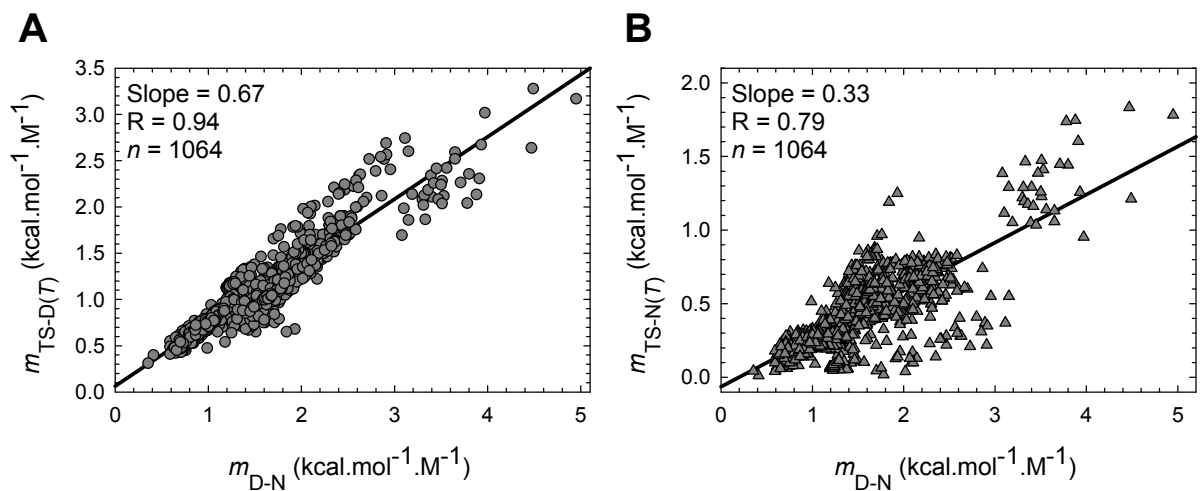

**Figure 9−figure supplement 1.**

**Experimentally determined mutation-induced variation in $m_{TS-D(T)}$ and $m_{TS-N(T)}$ with $m_{D-N}$ for a total of 1064 mutants from 31 two-state systems.**

When the raw data is not classified (as in **Figure 9**), approximately 67% of a unit change in $m_{D-N}$ is manifest in $m_{TS-D(T)}$ while the rest appears in $m_{TS-N(T)}$ with the sum of their slopes being unity. The slopes and the correlation coefficients of the linear regression are shown as insets.

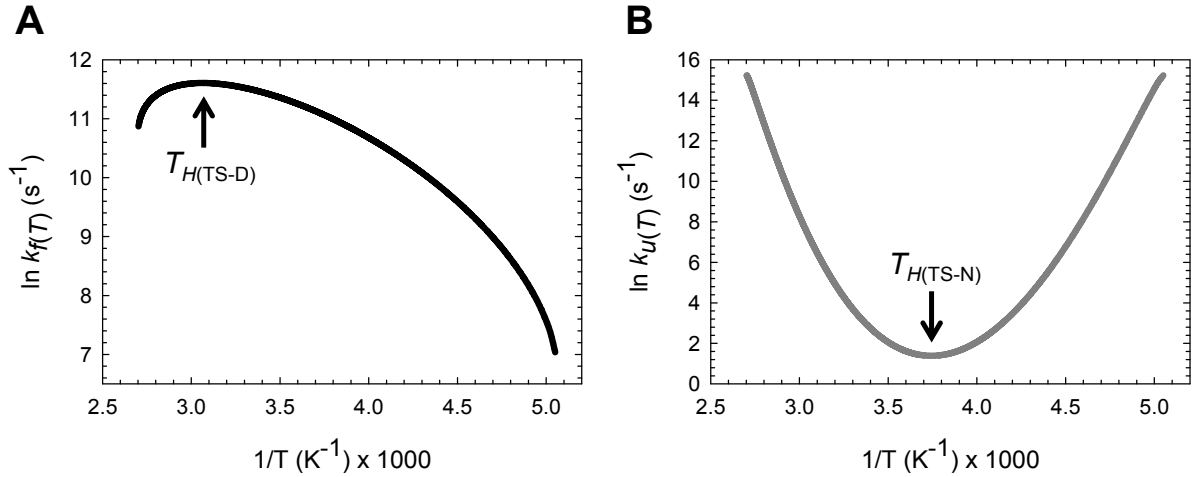

**Figure 10.**

**Arrhenius plots for the wild type BdpA Y15W.**

**(A)** Temperature-dependence of $k_{f(T)}$ (range 198 to 370 K) according to Eq. (35); $k_{f(T)}$ is a maximum and $\Delta H_{TS-D(T)} = 0$ at $T_{H(TS-D)} = 326.3$ K. The slope of this curve is given by $-\Delta H_{TS-D(T)}/R$ and is zero at $T_{H(TS-D)}$. **(B)** Temperature-dependence of $k_{u(T)}$ according to Eq. (36); $k_{u(T)}$ is a minimum and $\Delta H_{TS-N(T)} = 0$ at $T_{H(TS-N)} = 267.2$ K. The slope of this curve is given by $-\Delta H_{TS-N(T)}/R$ and is zero at $T_{H(TS-N)}$. The steep increase in $k_{u(T)}$ at very low and high temperatures is due to the Gibbs barrier height for unfolding approaching zero (addressed in subsequent publications). These data were generated using the following parameters: $k^0 = 4206663$ s$^{-1}$, $\alpha = 1.683$ M$^2$.mol.kcal$^{-1}$, $\omega = 122.175$ M$^2$.mol.kcal$^{-1}$, $m_{D-N} = 1.3936$ kcal.mol$^{-1}$.M$^{-1}$, $T_m = 350.3$ K, $\Delta H_{D-N(Tm)} = 50.85$ kcal.mol$^{-1}$ and $\Delta C_{pD-N} = 644$ cal.mol$^{-1}$.K$^{-1}$.[96] The values of the prefactor, the spring constants and $m_{D-N}$ were extracted from the chevron of BdpA Y15W at 298.16 K, and the data required to generate the chevron were taken from Table 7 of Sato and Fersht, 2007 (see **Methods**).[58] The force constants, prefactor, $m_{D-N}$ and $\Delta C_{pD-N}$ are temperature-invariant.

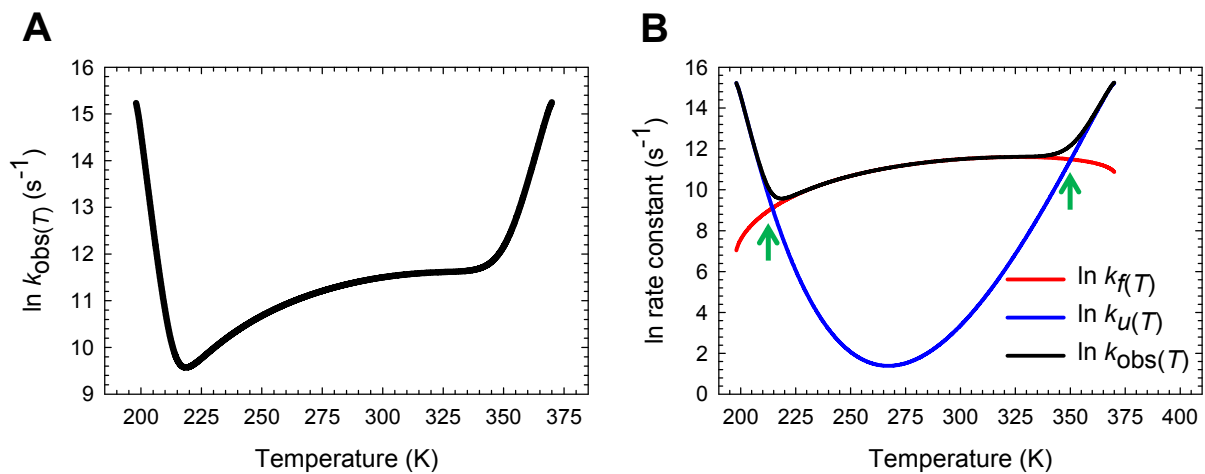

**Figure 10−figure supplement 1.**

**Temperature-dependence of the natural logarithm of the observed rate constant for BdpA Y15W.**

**(A)** Temperature-dependence of the natural logarithm of $k_{obs(T)}$ according to Eq. (37). **(B)** An overlay of the natural logarithm of $k_{f(T)}$, $k_{u(T)}$ and $k_{obs(T)}$. The steep increase in $k_{u(T)}$ and $k_{obs(T)}$ at very high and low temperatures is due to the Gibbs barrier height for unfolding approaching zero. The green pointers indicate the midpoints of cold ($T_c$ = 214.4 K) and heat denaturation ($T_m$ = 350.3 K) wherein $k_{f(T)} = k_{u(T)}$ (see Eq. (A26)). The slopes of the red and blue curves are given by $\Delta H_{TS\text{-}D(T)}/RT^2$ and $\Delta H_{TS\text{-}N(T)}/RT^2$, respectively, and are zero at $T_{H(TS\text{-}D)}$ and $T_{H(TS\text{-}N)}$, respectively.

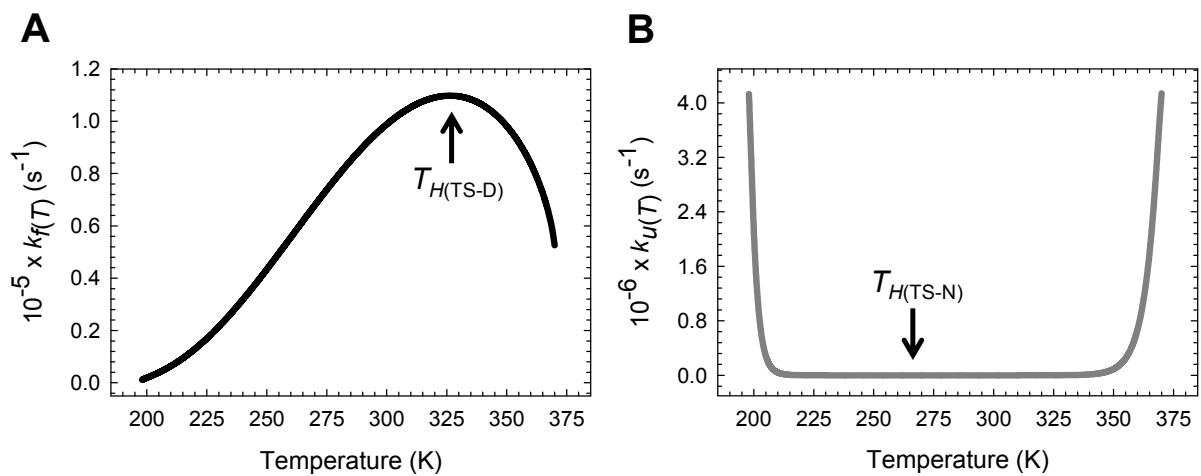

**Figure 10−figure supplement 2.**

**Temperature-dependence of $k_{f(T)}$ and $k_{u(T)}$ for BdpA Y15W on a linear scale.**

**(A)** The rate constant for folding is a maximum and $\Delta H_{TS\text{-}D(T)} = 0$ at $T_{H(TS\text{-}D)} = 326.3$ K. The slope of this curve is given by $k_{f(T)} \Delta H_{TS\text{-}D(T)} / RT^2$. **(B)** $k_{u(T)}$ is a minimum and $\Delta H_{TS\text{-}N(T)} = 0$ at $T_{H(TS\text{-}N)} = 267.2$ K. The slope of this curve is given by $k_{u(T)} \Delta H_{TS\text{-}N(T)} / RT^2$. The minimum of $k_{u(T)}$ is not apparent on a linear scale.

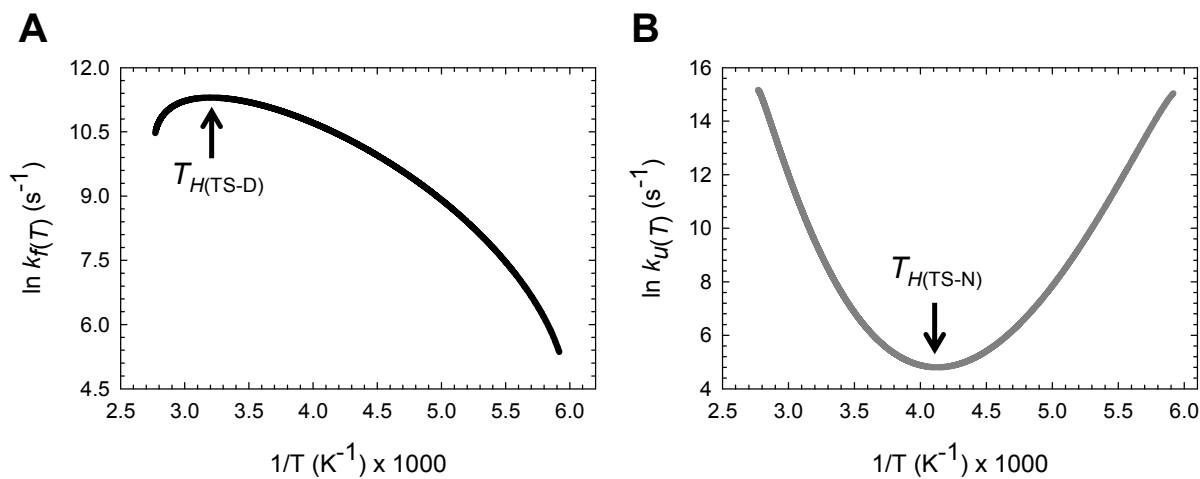

**Figure 11.**

**Arrhenius plots for the wild type BBL H142W**

**(A)** Temperature-dependence of $k_{f(T)}$ (range 168 to 362 K) according to Eq. (35); $k_{f(T)}$ is a maximum and $\Delta H_{\text{TS-D}(T)} = 0$ at $T_{H(\text{TS-D})} = 312.5$ K. The slope of this curve is given by $-\Delta H_{\text{TS-D}(T)}/R$ and is zero at $T_{H(\text{TS-D})}$. **(B)** Temperature-dependence of $k_{u(T)}$ according to Eq. (36); $k_{u(T)}$ is a minimum and $\Delta H_{\text{TS-N}(T)} = 0$ at $T_{H(\text{TS-N})} = 242.5$ K. The slope of this curve is given by $-\Delta H_{\text{TS-N}(T)}/R$ and is zero at $T_{H(\text{TS-N})}$. The steep increase in $k_{u(T)}$ at very low and high temperatures is due to the Gibbs barrier height for unfolding approaching zero. These data were generated using the following parameters: $k^0 = 3896195$ s$^{-1}$, $\alpha = 7.182$ M$^2$.mol.kcal$^{-1}$, $\omega = 283.793$ M$^2$.mol.kcal$^{-1}$, $m_{\text{D-N}} = 0.69$ kcal.mol$^{-1}$.M$^{-1}$, $T_m = 327.3$ K, $\Delta H_{\text{D-N}(T_m)} = 27$ kcal.mol$^{-1}$ and $\Delta C_{p\text{D-N}} = 350$ cal.mol$^{-1}$.K$^{-1}$.[59] The values of the prefactor, the spring constants and $m_{\text{D-N}}$ were extracted from the chevron of BBL H142W at 283 K, and the data required to generate the chevron were taken from Table 3 of Neuweiler et al., 2009 (see **Methods**).[59] The force constants, prefactor, $m_{\text{D-N}}$ and $\Delta C_{p\text{D-N}}$ are temperature-invariant.

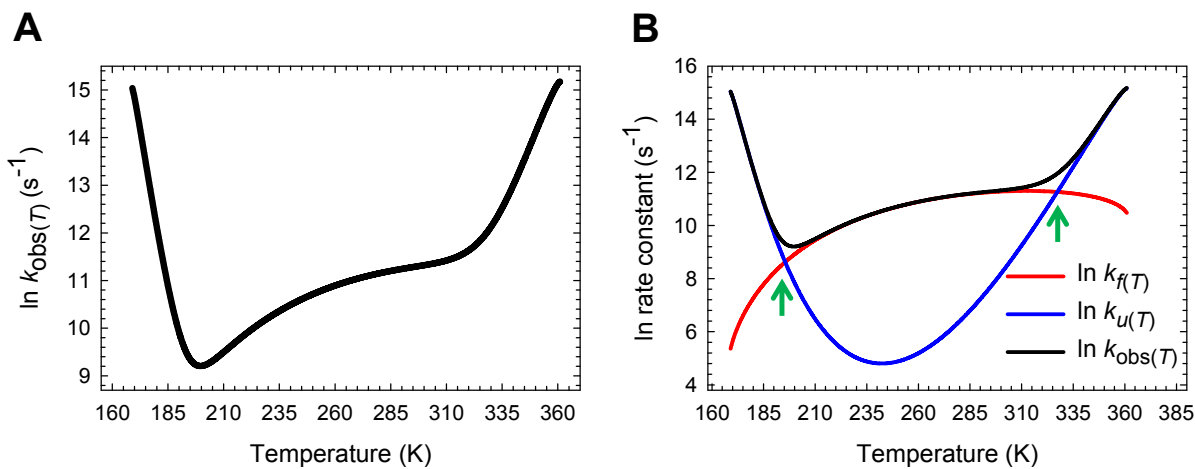

**Figure 11−figure supplement 1.**

**Temperature-dependence of the natural logarithm of the observed rate constant for BBL H142W.**

(A) Temperature-dependence of the natural logarithm of $k_{obs(T)}$ according to Eq. (37). (B) An overlay of the natural logarithm of $k_{f(T)}$, $k_{u(T)}$ and $k_{obs(T)}$. The steep increase in $k_{u(T)}$ and $k_{obs(T)}$ at very low and high temperatures is due to the Gibbs barrier height for unfolding approaching zero. The green pointers indicate $T_c$ (195.6 K) and $T_m$ (327.3 K) where $k_{f(T)} = k_{u(T)}$ (see Eq. (A26)). The slopes of the red and blue curves are given by $\Delta H_{TS-D(T)}/RT^2$ and $\Delta H_{TS-N(T)}/RT^2$, respectively, and are zero at $T_{H(TS-D)}$ and $T_{H(TS-N)}$, respectively.

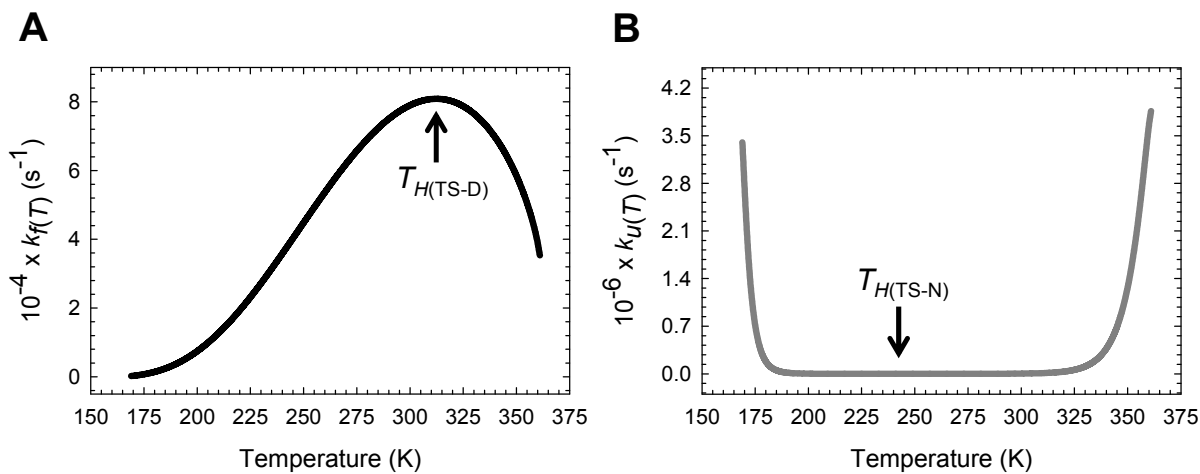

**Figure 11−figure supplement 2.**

**Temperature-dependence of $k_{f(T)}$ and $k_{u(T)}$ for BBL H142W on a linear scale.**

**(A)** $k_{f(T)}$ is a maximum and $\Delta H_{\text{TS-D}(T)} = 0$ at $T_{H(\text{TS-D})} = 312.5$ K. The slope of this curve is given by $k_{f(T)} \Delta H_{\text{TS-D}(T)} / RT^2$. **(B)** $k_{u(T)}$ is a minimum and $\Delta H_{\text{TS-N}(T)} = 0$ at $T_{H(\text{TS-N})} = 242.5$ K. The slope of this curve is given by $k_{u(T)} \Delta H_{\text{TS-N}(T)} / RT^2$. The minimum of $k_{u(T)}$ is not apparent on a linear scale.

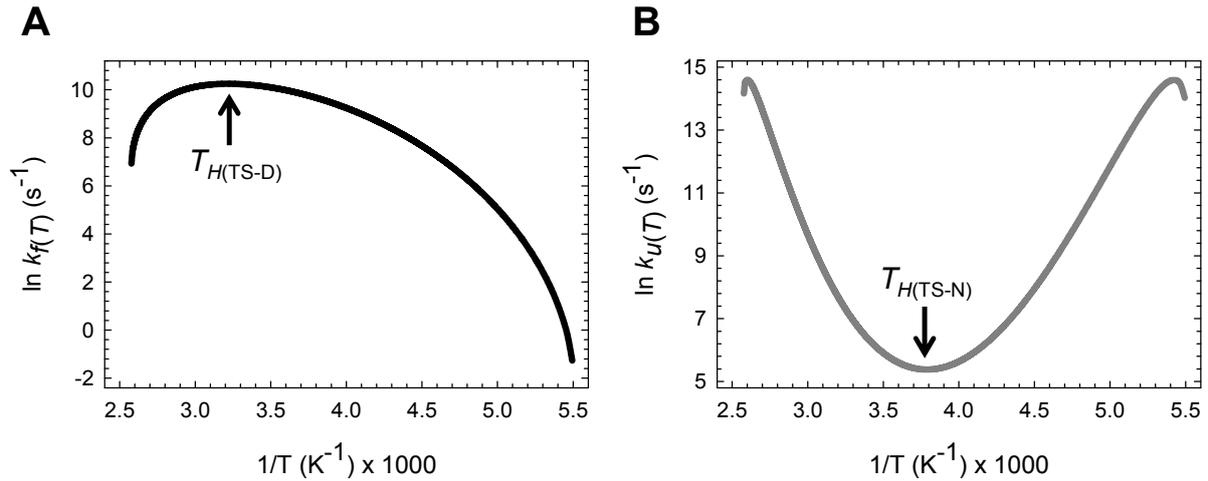

**Figure 12.**

**Arrhenius plots for the wild type FBP28 WW**

**(A)** Temperature-dependence of $k_{f(T)}$ (range 182 to 388 K) according to Eq. (35); $k_{f(T)}$ is a maximum and $\Delta H_{\text{TS-D}(T)} = 0$ at $T_{H(\text{TS-D})} = 311.4$ K. The slope of this curve is given by $-\Delta H_{\text{TS-D}(T)}/R$ and is zero at $T_{H(\text{TS-D})}$. **(B)** Temperature-dependence of $k_{u(T)}$ according to Eq. (36); $k_{u(T)}$ is a minimum and $\Delta H_{\text{TS-N}(T)} = 0$ at $T_{H(\text{TS-N})} = 264.3$ K. The slope of this curve is given by $-\Delta H_{\text{TS-N}(T)}/R$ and is zero at $T_{H(\text{TS-N})}$. The steep increase in $k_{u(T)}$ at very low and high temperatures, and its eventual saturation followed by rate-inversion is a consequence of the conventional barrier-limited unfolding going through barrierless regime and once again becoming barrier-limited (*Marcus-inverted-region*). These data were generated using the following parameters: $k^0 = 2180965$ s$^{-1}$, $\alpha = 7.594$ M$^2$.mol.kcal$^{-1}$, $\omega = 85.595$ M$^2$.mol.kcal$^{-1}$, $m_{\text{D-N}} = 0.82$ kcal.mol$^{-1}$.M$^{-1}$, $T_m = 337.2$ K, $\Delta H_{\text{D-N}(T_m)} = 26.9$ kcal.mol$^{-1}$ and $\Delta C_{p\text{D-N}} = 417$ cal.mol$^{-1}$.K$^{-1}$.[60] The values of the prefactor, the spring constants and $m_{\text{D-N}}$ were extracted from the chevron of FBP28 WW at 283.16 K, and the data required to generate the chevron were taken from Table 4 of Petrovich et al., 2006 (see **Methods**).[60] The force constants, prefactor, $m_{\text{D-N}}$ and $\Delta C_{p\text{D-N}}$ are temperature-invariant.

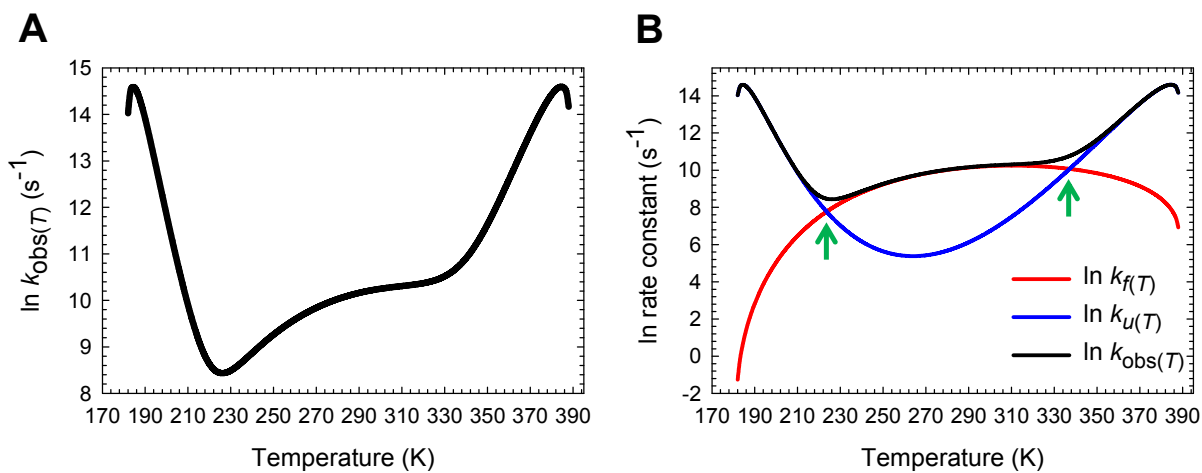

**Figure 12−figure supplement 1.**

**Temperature-dependence of the natural logarithm of the observed rate constant for FBP28 WW.**

**(A)** Temperature-dependence of the natural logarithm of $k_{obs(T)}$ according to Eq. (37). **(B)** An overlay of the natural logarithm of $k_{f(T)}$, $k_{u(T)}$ and $k_{obs(T)}$. The steep increase in $k_{u(T)}$ and $k_{obs(T)}$ at very low and high temperatures, and their eventual saturation followed by inversion is a consequence of the conventional barrier-limited unfolding going through barrierless regime and once again becoming barrier-limited (*inverted-region*). The green pointers indicate $T_c$ (223.6 K) and $T_m$ (337.2 K) where $k_{f(T)} = k_{u(T)}$ (see Eq. (A26)). The slopes of the red and blue curves are given by $\Delta H_{TS-D(T)}/RT^2$ and $\Delta H_{TS-N(T)}/RT^2$, respectively, and are zero at $T_{H(TS-D)}$ and $T_{H(TS-N)}$, respectively.

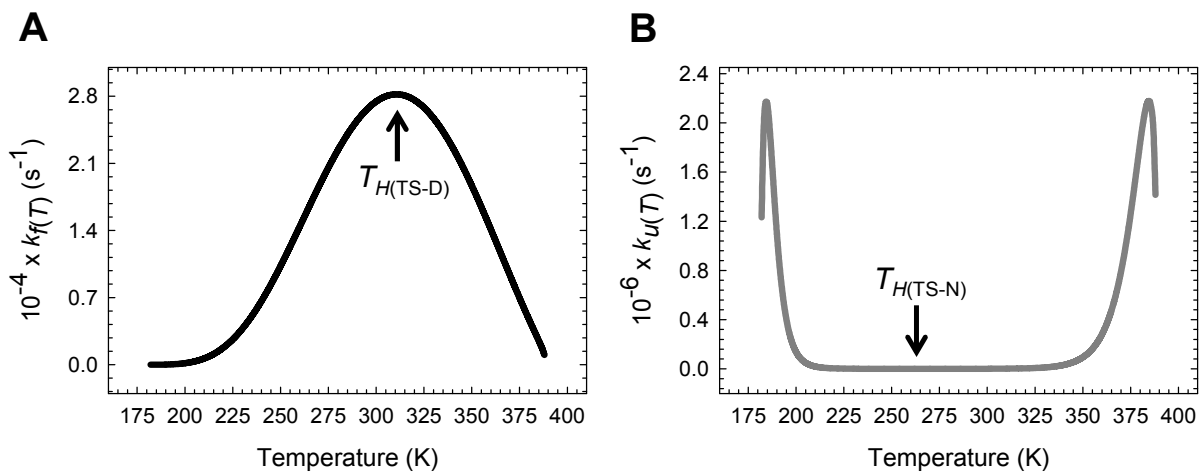

**Figure 12−figure supplement 2.**

**Temperature-dependence of $k_{f(T)}$ and $k_{u(T)}$ for FBP28 WW on a linear scale.**

**(A)** $k_{f(T)}$ is a maximum and $\Delta H_{TS\text{-}D(T)} = 0$ at $T_{H(TS\text{-}D)} = 311.4$ K. The slope of this curve is given by $k_{f(T)} \Delta H_{TS\text{-}D(T)} / RT^2$. **(B)** $k_{u(T)}$ is a minimum and $\Delta H_{TS\text{-}N(T)} = 0$ at $T_{H(TS\text{-}N)} = 264.3$ K. The slope of this curve is given by $k_{u(T)} \Delta H_{TS\text{-}N(T)} / RT^2$. The minimum of $k_{u(T)}$ is not apparent on a linear scale. The steep increase in $k_{u(T)}$ at very low and high temperatures, and their eventual saturation followed by inversion is a consequence of the conventional barrier-limited unfolding going through barrierless regime and once again becoming barrier-limited (*inverted-region*).

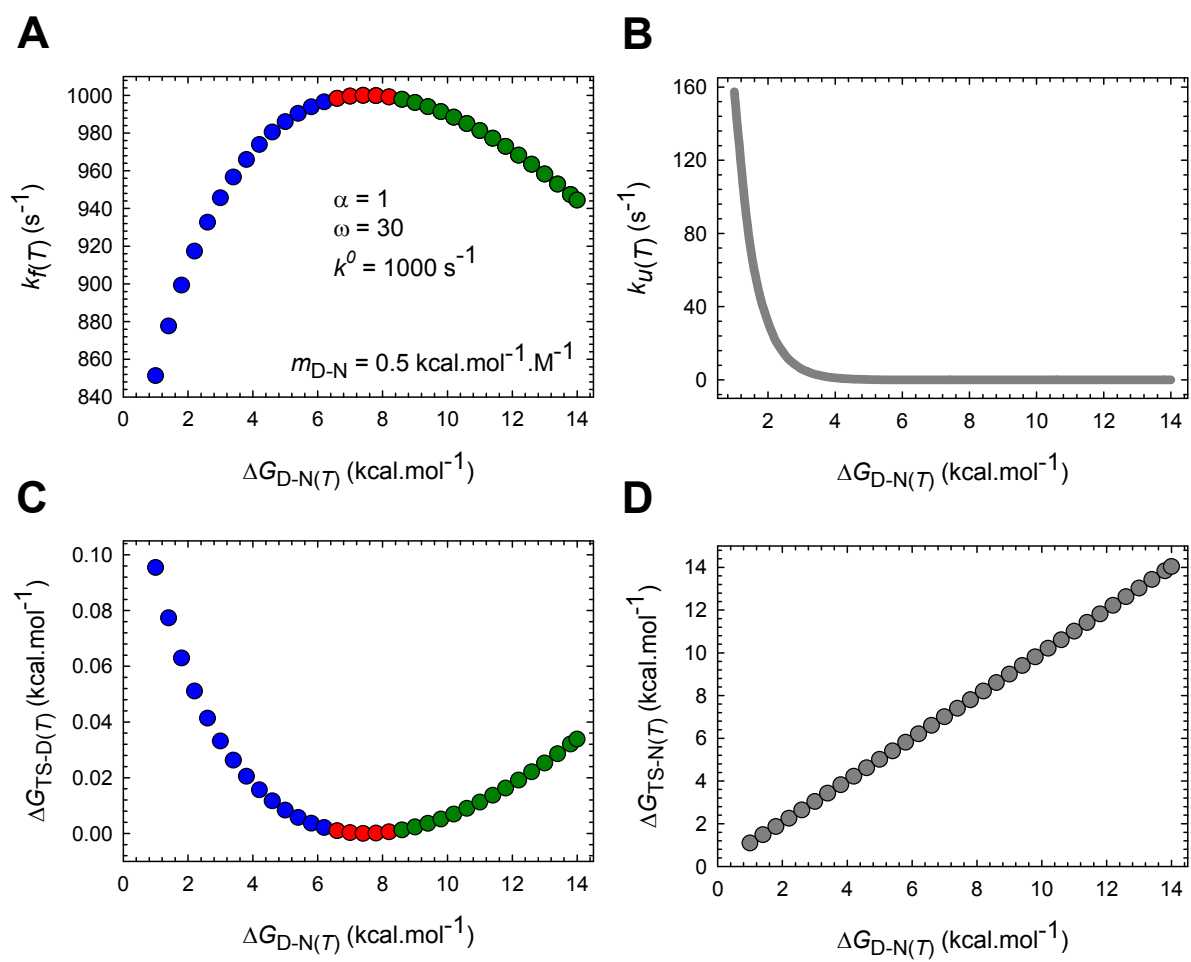

**Figure 13.**

**Conventional barrier-limited, barrierless and Marcus-inverted regimes for the folding reaction of a hypothetical two-state system according to Eqs. (11) and (12).**

The parameters used to simulate these curves are given in the top left panel ($T = 298.16$ K). The conventional barrier-limited, barrierless, and inverted regimes for folding are shown in blue, red, and green, respectively. When $m_{D-N}$ is very small, an increase in $\Delta G_{D-N(T)}$ can eventually lead to barrierless and inverted behaviour. **(A)** $k_{f(T)}$ increases with an increase in $\Delta G_{D-N(T)}$ and is a maximum ($k_{f(T)} = k^0$) at ~7 kcal.mol$^{-1}$; and any further increase in $\Delta G_{D-N(T)}$ beyond this point leads to a decrease in $k_{f(T)}$. **(B)** Exponential decrease in $k_{u(T)}$ with $\Delta G_{D-N(T)}$. **(C)** $\Delta G_{TS-D(T)}$ decreases with an increase in $\Delta G_{D-N(T)}$, is zero at ~7 kcal.mol$^{-1}$ and increases further on. **(D)** In contrast to $\Delta G_{TS-D(T)}$, $\Delta G_{TS-N(T)}$ increases in almost a linear fashion with stability.